\documentclass[12pt]{article}
\usepackage{amsmath}
\usepackage{amssymb}

\newtheorem{theorem}{Theorem}
\newtheorem{acknowledgement}[theorem]{Acknowledgement}

\begin{document}

\title{\textbf{Fair Division with money and prices:}\\
\textbf{Bid \& Sell versus Divide \& Choose}}
\author{Anna Bogomolnaia\thanks{%
University of Glasgow, U. K., and CNRS, France} and Herv\'{e} Moulin \thanks{%
University of Glasgow, U. K.}}
\date{ April 2025}
\maketitle

\begin{abstract}
We divide efficiently a pile of indivisible goods in common property, using
cash transfers to ensure fairness among agents with utility linear in money.
We compare three cognitively feasible and privacy preserving division rules
in terms of the guarantees (worst case utility) they offer to the
participants.

In the first version of Divide \& Choose to n agents, they bid for the role
of Divider then everyone bids on the shares of the Divider's partition. In
the second version each agent announces a partition and they all bid to
select the most efficient one.

In the Bid \& Sell rule the agents bid for the role of Seller: with two
agents the \textit{smallest} bid defines the Seller who then charges any
price constrained only by her winning bid.

Both rules reward subadditive utilities and penalise superadditive ones, and
B\&S more so than both D\&C-s. B\&S is also better placed to collect a
larger share of the surplus when agents play safe.


\begin{acknowledgement}
Moulin's research was supported in part by a Blaise Pascal Chair of the
Region Ile-de-France, 2020-21. The critical and helpful comments of two
referees are gratefully acknowledged.
\end{acknowledgement}
\end{abstract}

\section{Introduction}

The fair allocation of indivisible objects is greatly facilitated if the
agents who get few good objects or many bad ones accept compensations in
cash or any other transferable and divisible commodity (workload, stocks,
caviar, bitcoin). Examples of this common practice include the classic rent
division problem (\cite{GMPZ}, Spliddit.org), the dissolution of a
partnership (\cite{CGK}, the Texas Shoot Out clause to terminate a joint
venture\footnote{%
Both parties submit sealed bids and the party who makes the higher bid buys
the company at that price.}), and the NIMBY problem (the allocation of a
noxious facility between several communities \cite{KK}).

The familiar assumption that utilities are quasi-linear -- each agent can
attach to each bundle of objects a personal \textquotedblleft
price\textquotedblright\ and switching from one bundle to another is exactly
compensated by the difference in their prices -- yields a versatile fair
division model that the economic literature, so far, discussed with any
depth only in the special case of the \textit{assignment }problem where each
of the $n$ agents must receive at most one object (references in section 2).

We discuss the fair division of a finite number of indivisible \textit{goods}
(freely disposable objects) and money. Utilities are weakly increasing over
subsets of goods but externalities across goods are arbitrarily complex,
exactly like in the combinatorial auction problem (\cite{CSS}): if we
distribute the $m$ goods in a set $A$, a full description of an agent's
utility measured in money is a vector of dimension $2^{|A|}-1$.

We focus on two division rules, dubbed \textit{Divide and Choose} and 
\textit{Bid and Sell}, in which the message sent by each participant is of
much smaller dimension than $2^{|A|}$: for the former\ it is a single
partition of the objects and/or a set of transfers equalising one's utility
between the components of that partition; for the latter a single bid
followed by either selecting a price for each object or choosing to purchase
some goods at a given price vector. Of course computing my optimal safe
message in either rule relies on my entire utility functions, just like in
the auction context. But the information exchanged when playing the rules
remains cognitively simple, a critical requirement for their applicability
(as argued in \cite{PR}). Privacy protection is a \textquotedblleft
dual\textquotedblright\ argument against eliciting a full report, even when
the number of objects is small: revealing little of my preferences is an
advantage in subsequent bargaining interactions.

Consider the \textquotedblleft naive\textquotedblright\ division rule known
as \textit{Multi Auction} (MA): each agent $i$ places a bid $\beta _{ia}$
for each good $a$, the highest bidder $i^{\ast }$ on $a$ gets this object
and pays $\frac{1}{n}\beta _{i^{\ast }a}$ to each of the $n-1$ other agents.
Although MA is compelling if all utilities over the objects are additive, in
our much more general domain of utilities its performance is very poor: this
point is the object of section 11.1 of the Appendix.

Our two division rules of interest behave much better than MA. The first one
adapts to our context with cash transfers and any number $n$ of agents the
time honored \textit{Divide and Choose} (\textbf{D\&C}$^{1}$) method: a
round of bids determines the Divider agent, who picks a partition where each
lot contains some objects (possibly none) and some cash transfer (possibly
zero), after which each Chooser places bids summing to zero on the different
lots. We also discuss in section 6 a similar but more efficient version
denoted D\&C$^{2}$.

The second rule is the new \textit{Bid and Sell }rule (\textbf{B\&S}) where
each agent can have a role as Seller or Buyer. In the two agent case they
bid first to assign these roles, and a bid is interpreted as the price the
Seller can charge for \textit{all} the goods. The agent with the \textit{%
smallest} bid $x$ takes that role. The Seller then chooses a price for every
good so that their sum is $x$ and the Buyer can buy at those prices any
subset of goods, possibly all or none. The remaining goods go the the
Seller, along with the cash from the Buyer's purchase.

We compare the performance of our two rules mostly in terms of the \textit{%
ex ante guarantees} each agent secures by sending a \textit{safe} message. A
message by agent $i$ is safe if it maximises $i$'s worst case utility when
this agent only knows the number of other agents but not their utility
functions. Ensuring a high guarantee to each participant is the main
interpretation of ex ante fairness, pioneered in Steinhaus' work on cake
cutting (\cite{St1}, \cite{St2}). As in that model, a natural guarantee is $%
\frac{1}{n}u_{i}(A)$ for agent $i$ with utility $u_{i}$, that we call agent $%
i$'s \textit{Proportional Share} (PS). But unlike in that model, this
guarantee is not unique, and does not follow when we use one of the D\&C and
B\&C rules.

We argue that the PS guarantee, as the definition of ex ante fairness, is
much too coarse in our rich domain of division problems: we want instead to
reward agents with subadditive utilities and penalise those with
superadditive utilities.\smallskip

\textbf{Example 0}\textit{. We divide }$m\geq 2$\textit{\ identical goods
between two agents Frugal (female) and Greedy (male) with the following
utilities }%
\begin{equation*}
u_{F}(S)=1\text{ for all }S,\varnothing \neq S\subseteq A\text{ ; }%
u_{F}(\varnothing )=0
\end{equation*}%
\begin{equation*}
u_{G}(S)=0\text{ for all }S,\varnothing \subseteq S\varsubsetneq A\text{ ; }%
u_{G}(A)=1
\end{equation*}%
\textit{Frugal is content with any single good -- her utility is maximally
sub-additive -- while Greedy needs all goods to derive any utility -- his
utility is maximally super-additive.\smallskip }

We submit that it is not fair to offer ex ante the same PS guarantee $\frac{1%
}{2}$ to Frugal and Greedy. Under the veil of ignorance where we (as
impartial observer) don't know person $X$ who will share the goods with
Frugal, we should take into account that together Frugal and $X$ can produce
at least as much utility surplus -- and typically much more -- than if $X$
is paired with Greedy. The guarantee $\frac{1}{n}u_{i}(A)$ ignores this fact.

The \textit{Responsiveness} property says that we should guarantee \textit{%
strictly} more than her PS to Frugal, which implies that Greedy is
guaranteed strictly less than his PS (because the sum of utils of F and G is
1 for any division of the goods and cash). The \textit{Positivity }property,
by contrast, protects Greedy: it requires to give him \textit{some} positive
guarantee because his equal rights to the goods should amount to something
regardless of his uncompromising utility.

We compute first the guarantee offered by D\&C to Greedy in Example 0. He
must choose his bid $x$ to perhaps become the Divider knowing that there is
one other bidder, but clueless about the utility -- and possible bids -- of
that agent. So he will compute the \textit{worst case} utility that this bid
can get him.

We write $(S,t)$ for a share with the subset $S$ of goods and the cash
transfer $t$ (of arbitrary sign). If $x$ is the winning bid (that he first
pays to the other agent) his safe move as the Divider is to offer Chooser a
choice between the share $(A,-\frac{1}{2})$ (pay me $\frac{1}{2}$ and keep
all the goods) and $(\varnothing ,\frac{1}{2})$ (give me all of $A$ and I
will pay you $\frac{1}{2}$). In this way Greedy's utility from his
allocation is for sure $\frac{1}{2}$, and his net utility is $\frac{1}{2}-x$%
. If $x$ is the losing bid, he receives first at least $x$ from the winner
(whose bid is no less than $x$) then in the worst case faces a choice
between two allocations $(S,0)$ and $(A\diagdown S,0)$ where both shares are
non empty so that both allocations are worth zero to Greedy. \textit{It
happens here that Frugal will actually propose such a partition to optimise
her worst case. But Greedy's worst case analysis uses no such information:
he sees that in any other choice between }$(S,t)$\textit{\ and }$(A\diagdown
S,-t)$\textit{, for any subset }$S$\textit{\ and cash transfer }$t$\textit{,
he has a positive utility for at least one of the two allocations.}

Greedy's worst case utility if $x$ loses and he Chooses is $x$. Not knowing
if he wins or loses his worst net utility is the smallest of $x$ and $\frac{1%
}{2}-x$, which is largest for $x=\frac{1}{4}$ and guarantees him a gain of $%
\frac{1}{4}$. Any other bid than $\frac{1}{4}$ \textit{may} result in a
smaller gain.

Turning to Frugal, we compute first her worst possible utility for each of
the two possible roles after the bidding. As the Divider she secures the
utility of $1$ by offering a choice between $(S,0)$ and $(A\diagdown S,0)$
(where $S$ and $A\diagdown S$ are both non empty). \textit{As Chooser she
guarantees a net gain of }$\frac{1}{2}$\textit{\ for any choice between }$%
(S,t)$\textit{\ and }$(A\diagdown S,-t)$: indeed if $S$ is neither $A$ nor $%
\varnothing $, one of the shares has a non negative transfer so it is worth
at least a utility of $1$; and if the choice is between $(A,-t)$ and $%
(\varnothing ,t)$ she guarantees $\max \{1-t,t\}$ which is at least $\frac{1%
}{2}$. So Frugal's bid of $x$ in the first round secures the utility $1-x$
if she wins and $\frac{1}{2}+x$ if she loses: the smallest of these two is $%
\frac{3}{4}$ for her safe bid $x=\frac{1}{4}$ (and strictly less for any
other bid). The D\&C rule guarantees to Frugal three times more utility than
to Greedy.

In the Bid and Sell rule, the difference between Frugal's and Greedy's
guarantees sensibly increases as the number $m$ of goods grows so the
contrast between their preferences increases. We check that Greedy's
guaranteed utility is now $\frac{1}{m+1}$ versus $\frac{m}{m+1}$ for Frugal.

If $x$ is Frugal's initial bid to become the Seller and she loses, it means
that the (unknown) other agent's bid is \textit{smaller}, and as Seller that
agent must offer at least one good for a price at most $\frac{1}{m}x$
therefore Frugal can guarantee the net utility $1-\frac{1}{m}x$ by buying
just one such good. If Frugal becomes the Seller with the bid $x$, she will
safely post the uniform price $\frac{1}{m}x$ for each good: her net utility
is $1$ if she sells nothing, $x$ if she sells all the goods, and more than $%
1 $ if she sells some but not all goods: she gets at least $\min \{x,1\}$.
Choosing now $x$ to maximise $\min \{1-\frac{1}{m}x,\min \{x,1\}\}$, Frugal
picks $x=\frac{m}{m+1}$ and secures the net utility $\frac{m}{m+1}$.

Next consider Greedy with the initial bid $y$. His safe price as the Seller
offering to an unknown Buyer is uniform at $\frac{1}{m}y$: he gets $\frac{1}{%
m}y$ by selling at least one good and $1$ by selling nothing, which
guarantees the utility $\min \{\frac{1}{m}y,1\}$. As the Buyer, he will pay
at most $y$ for buying all the goods, which guarantees the utility $1-y$.
His safe bid $y=\frac{m}{m+1}$ maximises $\min \{\min \{\frac{1}{m}%
y,1\},1-y\}$: it is the same as Frugal's safe bid but only guarantees the
utility $\frac{1}{m+1}$ to Greedy.

\paragraph{Contents}

After the literature review in section 2 and the basic definitions in
section 3, we define guarantees in section 4. There we also describe simple
auctions implementing the \textit{fixed partition }guarantees, a key
ingredient of the D\&C rule.

Section 5 introduces two critical utility levels: the \textit{MaxMin utility}
that an agent can secure as the Divider is an upper bound on \textit{all}
guarantees; the \textit{MinMax utility }that she can secure as a Chooser
against an adversarial Divider is a lower bound on all reasonable
guarantees: Proposition 1.

Section 6 defines two versions of the D\&C rule and computes their
(different) safe play and (identical) guarantees: Proposition 2. Section 7
does the same for the B\&S rule: Proposition 3.

In section 8 we compare the PS, D\&C and B\&S guarantees. They share several
regularity and monotonicity properties (Lemma 5) as well as computational
complexity. Relative to the benchmark PS, the range of the B\&S guarantee is
much larger than that of the D\&C one: Proposition 4. But the coarser
messages in the D\&C rule can have strongly unpalatable consequences:
Example 3. Finally we compute explicitly our guarantees when the $m$ goods
are identical and utility are convex or concave (Lemma 6) or dichotomous
(Lemma 7).

Section 9 evaluates some welfare consequences of implementing one of our
individual guarantees. Does it distribute at least the total utility at the
worst partition of the goods? Lemma 8 gives some partial answers and
formulates a conjecture. Proposition 5 shows that, if agent $1$'s marginal
utility for each good dominates that of every other agent, then the B\&S
safe play achieves full efficiency, i.e., gives all the goods to agent $1$,
whereas under D\&C all but a $\frac{1}{n}$-th share of the efficient surplus
can be lost.

The concluding section 10 includes reports on numerical experiments
comparing the efficiency of safe play for our two main rules. With B\&S the
expected surplus is at least 95\% of the efficient one, whether utilities
are both superadditive, both subadditive, or mixed. The performance of the
D\&C rule is significantly weaker.

The Appendix (section 11) gathers several important proofs.

\section{Relevant literature}

Allowing cash compensations to smooth out the indivisibility of objects has
been essentially ignored by the first four decades of the theoretical
literature on fair division, if we except the cogent discussion by Steinhaus
of what we call above the Multi Auction rule for additive utilities (\cite%
{St2} p. 317).

This changed with the microeconomic discussion of the assignment problem.
Each agent wants at most one object and utilities are increasing in money
but not necessarily quasi-linear; monetary compensations can restore
fairness interpreted as Envy Freeness and even a version of the competitive
equilibrium with equal incomes: \cite{Sv} \cite{ADG}. The quasi-linear case
of the model is discussed in \cite{A} selecting a canonical envy free
allocation, in \cite{CGK} for the dissolution of partnership, in \cite{KK}
for adressing the NIMBY problem, and currently implemented on the
user-friendly Spliddit platform \cite{GP}.

In the assignment problem ex ante fairness is captured by the \textit{%
unanimous utility}: the best equal utility in the hypothetical problem where
everyone else shares my preferences (\cite{Mo3}, \cite{TT}). This is
unambiguously the best possible guarantee and it is compatible with Envy
Freeness.

In our model the set of allocations and utilities are vastly more complex
than in an assignment problem and the unanimity utility -- that we call the
MaxMin utility -- is an upper bound on guarantees but not itself a feasible
guarantee. Our newfound critique of Envy Freeness (Remarks 2, 3 in section
4.3, 5 respectively) complements the normative objections developed in \cite%
{Mo3}.

The search for a practical and appealing guarantee started the mathematical
cake cutting literature (\cite{St1}, \cite{Kuh}) and is a prominent theme in
the vibrant 21st century algorithmic literature on fair division surveyed in 
\cite{Mo4}, \cite{Az} and \cite{W}. There the \textit{standard model} has
utilities additive over objects and no cash transfers or lotteries, so the
definition of a convincing guarantee is complicated by the presence of
\textquotedblleft un-smoothable\textquotedblright\ indivisibilities. Our
MaxMin and MinMax utilities are the counterpart of, respectively, the
influential MaxMinShare due to Budish \cite{Bud} and its dual MinMaxShare 
\cite{BoLe}.\footnote{%
Other definitions of guarantees are also discussed in the algorithmic
literature, e. g. \cite{BTF}, as are guarantees adjusted to the granularity
of the utilities in \cite{DH}.} The MaxMinShare is \textit{almost }a
feasible guarantee (it is not feasible in extremely rare configurations \cite%
{PW}) while the dual MinMaxShare is strongly unfeasible. On the contrary in
our model the profile of MinMax utilities is always feasible (Lemma 3
section 5) while the MaxMin profile is unfeasible; this holds as well when
we divide a non atomic cake and utilities are continuous but otherwise
arbitrary: see \cite{BM1}, \cite{AvKa}.

In the standard model Envy Freeness is not feasible and one way to relax the
EF requirement is to allow cash transfers provably small in a certain sense;
these can (equivalently) come as non negative subsidies from the manager's
pocket or as a set of balanced transfers bewteen agents. The initial
positive result by \cite{HS} is strenghtened in \cite{BDNSV}, see also \cite%
{CI}, \cite{Az1}.

In the first of our two $n$-person versions of the Divide \& Choose rule
(section 6) the participants bid first for the role of Divider, which is
similar to and inspired by the auction in \cite{C} and \cite{De} for
implementing the egalitarian-equivalent division rule to distribute
Arrow-Debreu commodities.

\section{Basic definitions and notation}

\paragraph{Objects and money}

The finite set $A$ with cardinality $m\geq 2$ and generic elements $%
a,b,\cdots $, contains the indivisible objects that must \textbf{all} be
distributed between the $n$ agents in the set $N$ with generic elements $%
i,j,\cdots $ and $n\geq 2$.

With the familiar notation $[n]=\{1,\cdots ,n\}$ a $n$-partition $\pi $ of $%
A $ is a list $\pi =\{S_{k}\}_{k\in \lbrack n]}$ of \textbf{possibly empty}
and pairwise disjoint subsets of $A$ such that $A={\large \cup }_{k\in
\lbrack n]}S_{k}$: up to $n-1$ shares can be empty. If the relevant variable
is unambiguous we write a partition simply as $\{S_{k}\}_{[n]}$.

The set of $n$-partitions is $\mathcal{P}(n;A)$ if the shares $S_{k}$ are
not assigned to specific agents, and $\mathcal{P}(N;A)$ if they are.

Money is available in unbounded quantities to perform balanced transfers
between agents $t=(t_{i})_{i\in N}$ that are balanced: $\sum_{N}t_{i}=0$.
The set of such transfers is $\mathcal{T}(N)$. An \textit{allocation} is a
pair $(\pi ,t)\in \mathcal{P}(N;A)\times \mathcal{T}(N)$.

\paragraph{Utilities}

Each agent $i$ is endowed with a quasi-linear utility $u_{i}\in 
\mathbb{R}
^{2^{A}}$ over shares, with the important normalisation $u_{i}(\varnothing
)=0$: her utility from the allocation $(\pi ,t)$ is $u_{i}(S_{i})+t_{i}$.
The marginal utility of object $a$ at $S\subseteq A$ for utility $u$ is $%
\partial _{a}u(S)=u(S\cup a)-u(S\diagdown a)$. We assume throughout the
paper that all objects are \textit{goods}: $\partial _{a}u_{i}(S)\geq 0$ for
all $S\subseteq A$; utility functions can be any (weakly) inclusion
increasing non negative function on $2^{A}$, and $\mathcal{M}^{+}$ is our
notation for this domain.

The utility $u$ is additive if for all $a\in A$ the marginal $\partial
_{a}u(S)=u_{a}$ is independent of $S$; in this case we write $%
u_{S}=\sum_{S}u_{a}$ instead of $u(S)$.

We often use the following \textit{cover} operation to generate examples in
the domain $\mathcal{M}^{+}$.\footnote{%
It corresponds to an XOR bid in Nisan's\ terminology of bidding languages (%
\cite{Ni}).} Fix a subset $\{S_{k};1\leq k\leq K\}$ of $2^{A}\diagdown
\varnothing $ and $K$ positive utilities $v_{k}$; the \textit{cover} of the
subset $\{(S_{k},v_{k})\}$ of $2^{A}\diagdown \varnothing \times 
\mathbb{R}
_{+}$ is the smallest utility $u$ in $\mathcal{M}^{+}$ such that $%
u(S_{k})=v_{k}$ for all $k$:%
\begin{equation*}
u(S)=\max_{k:S_{k}\subseteq S}v_{k}\text{ ; }u(S)=0\text{ if }%
S_{k}\nsubseteq S\text{ for all }k
\end{equation*}%
For instance the Greedy utility $u_{G}$ in section 1 is the cover of $%
\{(A,1)\}$ while the Frugal utility $u_{F}$ is the cover of $\{(a,1);a\in
A\} $.

We call $u\in \mathcal{M}^{+}$ \textit{subadditive} if $u(S)+u(T)\leq
u(S\cup T)$ for all disjoint $S,T$ in $A$, and \textit{superadditive} if the
opposite inequalities hold. We write $\mathcal{S}ub$ and $\mathcal{S}up$ the
corresponding subsets of $\mathcal{M}^{+}$; their intersection is the set $%
\mathcal{A}dd$ of additive utilities.

\paragraph{Efficiency}

A $N$-profile of utilities is $\overrightarrow{u}=(u_{i})_{N}\in (\mathcal{M}%
^{+})^{N}$ and if $\pi \in \mathcal{P}(N;A)$ we write $\overrightarrow{u}%
(\pi )=\sum_{N}u_{i}(S_{i})$. An important special case is agent $i$'s 
\textit{unanimity} profile where all agents have the same utility $u_{i}$
that we write $(\overset{n}{u_{i}})$, so that $(\overset{n}{u_{i}})(\pi
)=\sum_{[n]}u_{i}(S_{k})$.

The notation $\overset{q}{z}$ for the $q$-vector with $q$ identical
coordinates $z$ will be used repeatedly.

The \textit{efficient surplus} at profile $\overrightarrow{u}$ is $\mathcal{W%
}(\overrightarrow{u})=\max_{\pi \in \mathcal{P}(N;A)}\overrightarrow{u}(\pi
) $. Recall an easy but critical consequence of the quasi-linearity
assumption\textit{: the allocation} $(\pi ^{\ast },t)\in \mathcal{P}%
(N;A)\times \mathcal{T}(N)$ \textit{is efficient (Pareto optimal: PO) if and
only if }$\pi ^{\ast }$\textit{\ maximises }$\overrightarrow{u}(\pi )$%
\textit{\ over} $\mathcal{P}(N;A)$. Pareto optimality is independent of the
balanced cash transfers.

\paragraph{Implementation}

Given an arbitrary $n$-agent mechanism agent $i$'s strategy is \textit{safe}
if it delivers to $i$ the largest \textquotedblleft worst
case\textquotedblright\ utility against all other agents playing
adversarially against $i$ after seeing $i$'s strategy. That utility is the 
\textit{guarantee} offered by this mechanism to agent $i$: it only depends
upon the mechanism, agent $i$'s utility function, and the number of other
agents.

Several mechanisms can implement the same guarantee: an example is the two
versions of Divide \& Choose in section 6. When computing guarantees we
systematically omit many tie-breaking details from the description of rules,
and the reader will find it easy to check that they (the details) never
affect the guarantee they implement.

At a given profile of utilities, in \textit{any }Nash equilibrium of the
game induced by the mechanism each agent gets at least their guaranteed
utility (otherwise this agent agent would benefit from deviating to a safe
strategy). Therefore how close is the sum of individual guarantees to the
efficient maximum is an upper bound on the price of anarchy: the worst loss
of efficiency at any equilibrium.

\section{Guarantees, Positive and Responsive}

\textbf{Definition 1}\textit{:} \textit{An }$n$\textit{-person guarantee is
a mapping} $\mathcal{M}^{+}\ni u\rightarrow \Gamma _{n}(u)\in 
\mathbb{R}
_{+}$ \textit{such that\ }%
\begin{equation}
\sum_{N}\Gamma _{n}(u_{i})\leq \mathcal{W}(\overrightarrow{u})\text{ for all 
}\overrightarrow{u}\in (\mathcal{M}^{+})^{N}  \label{13}
\end{equation}%
\textit{The set of }$n$\textit{-guarantees on }$A$\textit{\ is written} $%
\mathcal{G}(A;n)$.\smallskip

By inequality (\ref{13}) it is feasible at any utility profile $%
\overrightarrow{u}$ to give to each agent $i$ a share of surplus weakly
larger than $\Gamma _{n}(u_{i})$.

Guarantees are \textit{anonymous} by construction: they do not discriminate
between agents on the basis of their name. The three guarantees getting most
of our attention, Proportional Share, Bid \& Sell and Divide \& Choose, are
also \textit{neutral}, i. e., oblivious to the name of the objects in $A$.
So these guarantees only depend upon the numbers of objects and agents, and
the utility function of the concerned agent.

To any partition $\pi =\{S_{k}\}_{k\in \lbrack n]}\in \mathcal{P}(n;A)$ we
associate the $\pi $\textit{-guarantee }denoted%
\begin{equation*}
\Gamma _{n}^{\pi }(u)=\frac{1}{n}(\overset{n}{u})(\pi )=\frac{1}{n}%
\sum_{[n]}u(S_{k})\text{ for all }u\in \mathcal{M}^{+}
\end{equation*}

We check that $\Gamma _{3}^{\pi }$ meets inequality (\ref{13}) at an
arbitrary profile $\overrightarrow{u}=(u_{1},u_{2},u_{3})$; for a general $n$
the argument is quite similar. By definition of the efficient surplus the
three sums%
\begin{equation*}
u_{1}(S_{1})+u_{2}(S_{2})+u_{3}(S_{3})\text{ ; }%
u_{1}(S_{2})+u_{2}(S_{3})+u_{3}(S_{1})\text{ ; }%
u_{1}(S_{3})+u_{2}(S_{1})+u_{3}(S_{2})
\end{equation*}%
are bounded above by $\mathcal{W}(\overrightarrow{u})$. Taking the average
of these three inequalities gives the desired one: $\Gamma _{3}^{\pi
}(u_{1})+\Gamma _{3}^{\pi }(u_{2})+\Gamma _{3}^{\pi }(u_{3})\leq \mathcal{W}(%
\overrightarrow{u})$.

We speak of a generic $\pi $-guarantee (when $\pi $ is not specified) as a 
\textit{fixed partition guarantee}. The fixed partition guarantee
corresponding to the bundling partition $\pi ^{PS}=\{A,\overset{n-1}{%
\varnothing }\}$ is the familiar \textit{Proportional Share }(PS) $\Gamma
_{n}^{PS}(u)=\frac{1}{n}u(A)$.

\subsection{Implementing the $\protect\pi $-guarantees}

The simple \textbf{Bundle Auction (BA)} implements $\Gamma _{n}^{PS}$. Each
agent $i$ submits a non negative bid $\beta _{i}$ that the rule \textit{%
interprets} as this agent's utility for the entire set $A$; (one of) the
highest bidder(s) $i^{\ast }$gets $A$ and pays $\frac{1}{n}\beta _{i^{\ast
}} $ to each of the $n-1$ other agents.

The only safe bid in BA is the truthful one $\beta _{i}=u_{i}(A)$: it
guarantees to agent $i$ her PS $\frac{1}{n}u_{i}(A)$ while any other bid
risks delivering a smaller benefit: this is clear for a winning overbid, and
for an underbid losing to a bid between $\beta _{i}$and $u_{i}(A)$.\footnote{%
The tie break rule is irrelevant. The safe strategy and guarantee do not
change if the winner only pays $\frac{1}{n}$-th of the second highest price
to each loser.}

We generalise BA to the $\pi $\textbf{-auction }implementing the $\pi $%
-guarantee $\Gamma _{n}^{\pi }$ for any partition $\pi $ of $A$\textbf{.}
Given $\pi =\{S_{k}\}_{[n]}$ and the set $N$, each agent $i$ reports a
vector $t^{i}=(t_{k}^{i})_{[n]}\in \mathcal{T}(n)$ of balanced transfers
over those shares. The mechanism \textit{interprets} $t^{i}$ as equalising
agent $i$'s utility accross the different shares:%
\begin{equation}
\text{for all }k,\ell \in \lbrack n]:u_{i}(S_{k})+t_{k}^{i}=u_{i}(S_{\ell
})+t_{\ell }^{i}=\Gamma _{n}^{\pi }(u_{i})  \label{40}
\end{equation}%
which reveals the utilities $u_{i}(S_{k})$ up to an additive constant.

An \textit{assignment} of $\pi $ is a bijection $\sigma $ of $N$ into $[n]$,
and their set is $\mathcal{C}$. An assignment $\sigma ^{\ast }$ is optimal
at $\overrightarrow{u}$ if it maximises $\sum_{N}u_{i\sigma (i)}$ over $%
\mathcal{C}$. If each utility $u_{i}$ meets equation (\ref{40}) this is the
same as minimising the \textquotedblleft slack\textquotedblright\ $\delta
(\sigma )=\sum_{N}t_{\sigma (i)}^{i}$ over $\mathcal{C}$.

Because each $t^{i}$ is balanced we have $\sum_{\mathcal{C}}\delta (\sigma
)=0$, therefore the minimal slack $\delta (\sigma ^{\ast })$ is negative or
zero. After each agent $j$ receives\ $t_{\sigma ^{\ast }(j)}^{j}$ (a cash
handout if $t_{\sigma ^{\ast }(j)}^{j}>0$, a tax if $t_{\sigma ^{\ast
}(j)}^{j}<0$) the remaining cash surplus $|\delta (\sigma ^{\ast })|$ is
divided equally between all agents. Agent $i$'s final allocation is $%
(S_{\sigma ^{\ast }(i)},t_{\sigma ^{\ast }(i)}^{i}+\frac{1}{n}|\delta
(\sigma ^{\ast })|)$ for which her utility is $\Gamma _{n}^{\pi }(u_{i})+%
\frac{1}{n}|\delta (\sigma ^{\ast })|$.

We illustrate the $\pi $\textit{-}guarantees and their implementation with a
three good, three agent example, on which we apply more concepts and results
until section 7.\smallskip

\textbf{Example 1 }\textit{Three agents }$X,Y,Z$\textit{\ share three goods }%
$a,b,c$\textit{\ and their utilities are}%
\begin{equation}
\begin{array}{cccccccc}
& \text{a} & \text{b} & \text{c} & \text{ab} & \text{ac} & \text{bc} & \text{%
abc} \\ 
\text{X} & 9 & 6 & 0 & 15 & 12 & 15 & 15 \\ 
\text{Y} & 15 & 15 & 15 & 15 & 18 & 18 & 18 \\ 
\text{Z} & 6 & 3 & 0 & 6 & 6 & 6 & 21%
\end{array}
\label{39}
\end{equation}

Note that Y's utility is almost Frugal, while Z's is somewhat Greedy.

Consider the partition $\pi =\{ac,b,\varnothing \}$ with corresponding
utilities $(12,6,0)$ for X. The report $t^{X}=(-6,0,+6)$ of balanced
transfers defined by (\ref{40}) is X's unique safe report securing the
utility $\Gamma _{3}^{\pi }(u_{X})=6$ for each of the three shares $%
(ab,-6),(b,0),(\varnothing ,+6)$. Lemma 1 below proves this for a general
problem.

Computing similarly the balanced transfers equalising Y's (resp. Z's)
utilities for the shares $(ab,t_{ac}),(b,t_{b}),(\varnothing ,t_{\varnothing
})$ gives:%
\begin{equation}
\begin{array}{cccc}
& t_{ac} & t_{b} & t_{\varnothing } \\ 
\text{X} & -6 & 0 & +6 \\ 
\text{Y} & -7 & -4 & +11 \\ 
\text{Z} & -3 & 0 & +3%
\end{array}
\label{41}
\end{equation}%
from which we get the individual guarantees%
\begin{equation}
(\Gamma _{3}^{\pi }(u_{X}),\Gamma _{3}^{\pi }(u_{Y}),\Gamma _{3}^{\pi
}(u_{Z}))=(6,11,3)  \label{43}
\end{equation}

Upon comparing in matrix (\ref{41}) the slack of the six assignments of the
shares $ac$, $b$ and $\varnothing $ to the agents X,Y and Z, we find that $%
\sigma ^{\ast }$ giving $ac$ to X, $b$ to Y and nothing to Z is efficient:
it generates the smallest slack $\delta (\sigma ^{\ast })=-6-4+3=-7$. Then
we rebate to each agent $\frac{1}{3}$ of $|\delta (\sigma ^{\ast })|$, that
is $2\frac{1}{3}$. The final allocation and utility profile are%
\begin{equation}
\text{X}:(ac,-3\frac{2}{3})\text{, Y}:(b,-1\frac{2}{3})\text{, Z}%
:(\varnothing ,5\frac{1}{3})  \label{42}
\end{equation}%
\begin{equation*}
(u_{X},u_{Y},u_{Z})=(8\frac{1}{3},13\frac{1}{3},5\frac{1}{3})
\end{equation*}

This is the profile of utilities when each agent reports safely (hence
truthfully). Here and in general this is much more than their guaranteed
utility. Indeed the $\pi $-auction implements the most efficient assignment
of $\pi $. In particular if all agents report safely (i. e., truthfully) the
final allocation will be efficient over all partitions if and only if $\pi $
happens to be an efficient partition. Agent $i$'s lower utility $\Gamma
_{3}^{\pi }(u_{i})$ is reached only when the other two agents report
\textquotedblleft adversarial\textquotedblright\ transfers resulting in a
null slack.\smallskip

\textbf{Lemma 1} \textit{The }$\pi $\textit{-auction implements the }$\pi $%
\textit{-guarantee, and the unique safe play is to report the transfers
equalising one's utility across the shares of }$\pi $\textit{\ (as in (\ref%
{40})).}

\textbf{Proof }We fix $t^{1}\in \mathcal{T}(n)$ and compute agent $1$'s
worst utility after reporting $t^{1}$.

Check first that any $\sigma $ in $\mathcal{C}$ can be selected as uniquely
optimal for some reports of the other agents. Suppose that all other agents $%
j$ report $t^{1}$ as well: then $\sum_{i\in N}t_{\tau (i)}^{1}=0$ for any
assignment $\tau $ so they are all equally optimal. For each agent $j\neq 1$%
, assigned $S_{\sigma (j)}$ by the given $\sigma $, we modify $j$'s report
as follows%
\begin{equation*}
t_{\sigma (j)}^{j}=t_{\sigma (j)}^{1}-\varepsilon \text{ ; }t_{\ell
}^{j}=t_{\ell }^{i}+\frac{1}{n-1}\varepsilon \text{ for all }\ell \neq
\sigma (j)
\end{equation*}%
indicating that $j$ likes the share $S_{\sigma (j)}$ relative to the other
shares $\frac{n}{n-1}\varepsilon $ more than $1$ does. The slack of
assignment $\sigma $ is now $\delta (\sigma )=\sum_{N}t_{\sigma
(i)}^{i}=-(n-1)\varepsilon $, smaller than for any other assignment in which
at least one corrective term is positive. So $\sigma $ is selected as
announced, and results in agent $1$'s final utility $u_{1}(S_{\sigma
(1)})+t_{\sigma (1)}^{1}+\frac{n-1}{n}\varepsilon $.

As $\sigma $ and $\varepsilon $ were arbitrary we see that $i$'s utility
could be as low as $\min_{k\in \lbrack n]}u_{i}(S_{k})+t_{k}^{i}$. The
unique choice of $t^{1}$ maximising the latter equalises $i$'s utility
across these shares as in (\ref{40}), and secures the utility $\frac{1}{n}%
\sum_{k\in \lbrack n]}u_{i}(S_{k})=\frac{1}{n}(\overset{n}{u_{i}})(\pi )$,
while any other report is unsafe. $\blacksquare $

\subsection{The averaging auction}

The set $\mathcal{G}(A;n)$ of $n$-guarantees is clearly convex.

For an arbitray finite set $\{\pi ^{r}\}$ of partitions in $\mathcal{P}(N;A)$
indexed by $r\in R$ we describe the canonical implementation of the average
guarantee $\frac{1}{|R|}\sum_{R}\Gamma _{n}^{r}$, which we call the \textbf{%
averaging-auction}. This is a key component of the second Divide \& Choose
rule in section 6.

Each agent $i$ reports balanced transfers $t^{i}=(t_{r}^{i})_{R}\in \mathcal{%
T}(R)$ over those guarantees, interpreted as equalising the utilities $%
\Gamma _{n}^{r}(u_{i})$:%
\begin{equation}
\text{for all }r,s\in R\text{: }\Gamma _{n}^{r}(u_{i})+t_{r}^{i}=\Gamma
_{n}^{s}(u_{i})+t_{s}^{i}=\frac{1}{|R|}\sum_{R}\Gamma _{n}^{r}(u_{i})
\label{16}
\end{equation}

Then we select a guarantee $\Gamma _{n}^{r^{\ast }}$ at which the sum of the
corresponding transfers is minimal:%
\begin{equation*}
r^{\ast }\in \arg \min_{R}\sum_{N}t_{r}^{i}=\arg \max_{R}\sum_{N}\Gamma
_{n}^{r}(u_{i})
\end{equation*}

Call $\theta (r)=\sum_{N}t_{r}^{i}$ the slack of partition $r$ in $R$ and
note that $\sum_{R}\theta (r)=0$ implies $\theta (r^{\ast })\leq 0$. We
divide the surplus $|\theta (r^{\ast })|$ equally and the net transfer to
agent $i$ is $t_{r^{\ast }}^{i}+\frac{1}{n}|\theta (r^{\ast })|$. Finally $%
\Gamma _{n}^{r^{\ast }}$ is implemented and agent $i$'s net utility is at
least $\Gamma _{n}^{r^{\ast }}(u_{i})+t_{r^{\ast }}^{i}+\frac{1}{n}|\theta
(r^{\ast })|$.

From equation (\ref{16}) and the fact that if we fix $u_{i}$ some choices
of\ the other agents' utilities generate the slack $\theta (r)=0$ for all $r$%
, we conclude that $i$'s guaranteed utility is exactly $\frac{1}{|R|}%
\sum_{R}\Gamma _{n}^{r}(u_{i})$ as desired.\smallskip

\textbf{Example 1} \textit{(continued) }We describe the implementation of
the average $\frac{1}{2}\Gamma _{3}^{PS}+\frac{1}{2}\Gamma _{3}^{\pi }$ where%
\textit{\ }$\pi =\{ac,b,\varnothing \}$\textit{\ as above.} From the earlier
computation of $\Gamma _{3}^{\pi }$ for this example, and $\Gamma
_{n}^{PS}(u_{i})=\frac{1}{3}u_{i}(A)$, we compute the two guarantees and
corresponding transfer vectors given by equation (\ref{16}):%
\begin{equation*}
\begin{array}{ccc}
& \Gamma _{3}^{PS}(u_{i}) & \Gamma _{3}^{\pi }(u_{i}) \\ 
\text{X} & 5 & 6 \\ 
\text{Y} & 6 & 11 \\ 
\text{Z} & 7 & 3%
\end{array}%
\Longrightarrow 
\begin{array}{ccc}
& t_{\Gamma _{3}^{PS}} & t_{\Gamma _{3}^{\pi }} \\ 
\text{X} & 0.5 & -0.5 \\ 
\text{Y} & +2.5 & -2.5 \\ 
\text{Z} & -2 & +2 \\ 
\theta (r) & +1 & -1%
\end{array}%
\end{equation*}

As $t_{\Gamma _{3}^{\pi }}<t_{\Gamma _{3}^{PS}}$ we see that the $\pi $%
-auction brings more surplus than the bundle auction. So X and Y compensate
Z as shown in the column $t_{\Gamma _{3}^{\pi }}$, and an equal share of the
slack, $\frac{1}{3}$, is rebated to everyone. Then we implement $\Gamma
_{3}^{\pi }$ and the final utilities are $(u_{X},u_{Y},u_{Z})=(5\frac{5}{6},8%
\frac{5}{6},5\frac{1}{3})$.\ Comparing with (\ref{43}) Z is much better off
than under $\Gamma _{3}^{\pi }$ whereas X,Y are worse off.\smallskip

\textbf{Lemma 2} \textit{The averaging-auction implements the average
guarantee }$\frac{1}{|R|}\sum_{R}\Gamma _{n}^{r}$\textit{. The unique safe
play is to report the transfers equalising one's utility across guarantees
(as in (\ref{16})).\smallskip }

The straightforward proof, similar to that of Lemma 1, is omitted.\smallskip

\textbf{Remark 1}\textit{\ }It is just as easy to implement any convex
combination of guarantees $\sum_{R}\lambda _{r}\Gamma _{n}^{r}$\ where each $%
\lambda _{r}$\ is positive and $\sum_{R}\lambda _{r}=1$. Each agent $i$\
reports a vector of $\lambda $-balanced transfers $t^{i}$, $\sum_{R}\lambda
_{r}t_{r}^{i}=0$, and the rule proceeds as before: it implements $\Gamma
_{n}^{r^{\ast }}$\ where $r^{\ast }$\ minimises $\sum_{N}t_{r}^{i}$\ so the
slack $\theta (r^{\ast })=\sum_{i\in N}t_{r^{\ast }}^{i}$\ is still non
positive and $i$\ receives $t_{r^{\ast }}^{i}+\frac{1}{n}|\theta (r^{\ast
})| $. The safe strategy is to choose $\lambda $-balanced transfers $t^{i}$\
equalising utilities as in (\ref{16}).

\subsection{Positivity and Responsiveness}

The next two properties generalise the argument developed in Example 0 in
the Introduction. Recall the Frugal utility, $u_{F}:u_{F}(S)=1$ for $S\neq
\varnothing $, and Greedy one, $u_{G}:u_{G}(S)\equiv 0$ for $S\neq A$, $%
u_{G}(A)=1$.\smallskip

\textbf{Definition 2} \textit{The }$n$\textit{-guarantee} $\Gamma _{n}\in 
\mathcal{G}(n,A)$ \textit{is}

\noindent \textit{Positive} \textit{if for all} $u\in \mathcal{M}%
^{+}:u(A)>0\Longrightarrow \Gamma _{n}(u)>0$

\noindent \textit{Responsive} \textit{if }$\Gamma _{n}(u_{F})>\frac{1}{n}%
>\Gamma _{n}(u_{G})$\smallskip

If Positivity fails at the utility $u$ of agent $i$, the goods are the
common property of all the agents and yet deliver no benefit to agent $i$ to
whom they are valuable: this normative position is untenable.

For Responsiveness we observe first that the inequality $\Gamma _{n}(u_{F})>%
\frac{1}{n}$ implies $\Gamma _{n}(u_{G})<\frac{1}{n}$, because the efficient
surplus is $1$ when $(n-1)$ Greedy agents share $A$ with a single Frugal
agent. So Responsiveness boils down to $\Gamma _{n}(u_{F})>\frac{1}{n}$.

We justify the latter inequality by comparing the contributions to the
efficient surplus of a Frugal versus a Greedy agent. Fix a $(n-1)$-profile $%
u_{-1}\in (\mathcal{M}^{+})^{n-1}$ and note that $\mathcal{W}%
(u_{F},u_{-1})\geq \mathcal{W}(u_{G},u_{-1})$. If this is an equality we
pick a partition $\pi =\{S_{i}\}_{N}$ efficient at $(u_{G},u_{-1})$ and we
have%
\begin{equation*}
u_{G}(S_{1})+\sum_{i\geq 2}u_{i}(S_{i})=u_{F}(S_{1})+\sum_{i\geq
2}u_{i}(S_{i})
\end{equation*}%
implying that $S_{1}$ is $\varnothing $ or $A$.

If $S_{1}=\varnothing $ both versions of agent 1 contribute nothing to the
efficient surplus, and if $S_{1}=A$ all $(n-1)$ other agents are equally
useless. Hence replacing a Greedy agent by a Frugal one always brings more
surplus if there is at least one efficient allocation of the goods where
Frugal shares the goods with the $(n-1)$ others, whoever they are.

Among the fixed partition guarantees, only $\Gamma _{n}^{PS}$ is Positive.
All fixed partition guarantees are Responsive, with the single exception of $%
\Gamma _{n}^{PS}$. Thus a convex mixture of $\Gamma _{n}^{PS}$ with any
other $\pi $-guarantees meets both properties.\smallskip

\textbf{Remark 2\ }The standard interpretation of ex post fairness in our
model is Envy Freeness (EF): the allocation $(\pi ,t)$\ is EF if $%
u_{i}(S_{i})+t_{i}\geq u_{i}(S_{j})+t_{j}$\ for all $i,j\in N$.
Surprisingly, Positivity and Responsiveness are not together compatible with
Envy Freeness! If the $n$-guarantee $\Gamma _{n}$\ in $\mathcal{M}^{+}$\ is
Positive and Responsive, then a rule implementing it cannot choose an
envy-free allocation at all utility profiles.

Proof by contradiction. We fix such a guarantee $\Gamma _{n}$ implemented by
a rule selecting at each utility profile an EF allocation. At the profile
with $(n-1)$ Greedy agents and a single Frugal we assume first that some
agent gets all of $A$, with identical value $1$ for everyone. By EF that
agent pays $\frac{1}{n}$ to everyone else and all end up with utility $\frac{%
1}{n}$: this contradicts Responsiveness for Frugal. If the goods are split
between at least two agents, by Positivity every Greedy one gets some
positive transfer, and by EF all get the same transfer $t$, so Frugal pays $%
(n-1)t$. But then Frugal envies at least one Greedy agent who gets some good.

\section{MaxMin and MinMax utilities}

The recent literature on fair division pays close attention to these two
canonical utility levels inspired by Divide \& Choose for cake-cutting, but
playing a role in many other models. Recall the notation $(\overset{n}{u})$
for the unanimity profile where all $n$ agents have utility $u$.\smallskip

\textbf{Definition 3} \textit{Fix }$A$\textit{,}$n$\textit{\ and} $u\in 
\mathcal{M}^{+}$.

\noindent $i)$ \textit{The MaxMin utility at }$u$\textit{\ is} $%
MaxMin_{n}(u)=\frac{1}{n}\max_{\pi \in \mathcal{P}(n;A)}(\overset{n}{u})(\pi
)$\textit{: the largest utility agent }$u$\textit{\ can secure by choosing
an (anonymous) allocation} $(\pi ,t)\in \mathcal{P}(n;A)\times \mathcal{T}%
(n) $ \textit{and eating his worst share }$(S_{k},t_{k})$ \textit{of that
allocation.}

\noindent $ii)$ \textit{The MinMax utility at }$u$\textit{\ is} $%
MinMax_{n}(u)=\frac{1}{n}\min_{\pi \in \mathcal{P}(n;A)}(\overset{n}{u})(\pi
)$\textit{: the largest utility agent }$u$\textit{\ can secure by picking
her best share in the worst possible (anonymous) allocation} $(\pi ,t)\in 
\mathcal{P}(n;A)\times \mathcal{T}(n)$.\smallskip

Given an $n$-partition $\pi =\{S_{k}\}_{[n]}$ of $A$, the $\pi $-auction
guarantees the utility $\frac{1}{n}(\overset{n}{u_{i}})(\pi )$ to each agent 
$i$ (Lemma 1) therefore $i$ reaches her $MaxMin$ utility if she can choose $%
\pi $, and at least her $MinMax$ one if the choice of $\pi $ is
adversarial.\smallskip

\textbf{Example 1}\textit{\ (continued)}

\noindent Consider agent X. The partition $\pi _{1}=\{bc,a,\varnothing \}$
gives $(\overset{n}{u_{X}})(\pi _{1})=24$, and every other partition gives
her less. By attaching balanced transfers to the shares agent X ensures that
all three shares are worth $\frac{24}{3}=8$, thus maximising her utility for
the worst share: $MaxMin_{3}(u_{X})=8$. For $MinMax_{3}(u_{X})$ note that
the three partitions $\{abc,\varnothing ,\varnothing \}$, $\{a,b,c\}$, $%
\{ab,c,\varnothing \}$ minimise $(\overset{n}{u_{X}})(\pi )$ at the level $%
15 $. The worst balanced transfers attached to any such partition make all
the shares worth $5$ to X and any other choice allows at least one share to
give X more utility: $MinMax_{3}(u_{X})=5$.

Similar computations for Y and Z give%
\begin{equation}
\begin{array}{ccc}
& MaxMin_{3} & MinMax_{3} \\ 
\text{X} & 8 & 5 \\ 
\text{Y} & 11 & 6 \\ 
\text{Z} & 7 & 2%
\end{array}
\label{38}
\end{equation}

\textbf{Lemma 3} \textit{In the domain} $\mathcal{M}^{+}$

\noindent $i)$ \textit{If }$A$\textit{\ contains at least two goods, the
mapping }$u\rightarrow MaxMin_{n}(u)$ \textit{is not a }$n$-\textit{%
guarantee (property (\ref{13}) fails)}\newline
\textit{but it is an upper bound for every guarantee }$\Gamma _{n}\in 
\mathcal{G}(A;n)$:%
\begin{equation*}
\Gamma _{n}(u)\leq MaxMin_{n}(u)\text{ for all }u\in \mathcal{M}^{+}
\end{equation*}%
$ii)$ \textit{The mapping} $u\rightarrow MinMax_{n}(u)$ \textit{is a }$n$%
\textit{-guarantee:} $MinMax_{n}(\cdot )\in \mathcal{G}(A;n)\smallskip $

\textbf{Proof} For $i)$ we fix an arbitrary guarantee $\Gamma _{n}$ and
utility $u$. Inequality (\ref{13}) at the unanimity profile $(\overset{n}{u}%
) $ is $n\Gamma _{n}(u)\leq \max_{\pi \in \mathcal{P}(n;A)}u(\pi )$ as
desired.

To check that $MaxMin$ is not a guarantee we have%
\begin{equation*}
MaxMin_{n}(u_{F})=\min \{1,\frac{m}{n}\}\text{ and }MaxMin_{n}(u_{G})=\frac{1%
}{n}
\end{equation*}%
because if $n\leq m$ Frugal can choose the partition with $n$ shares
containing a single object, but if $n>m$ she can only offer $m$ such shares.
The only valuable partition to Greedy bundles $A$ as a single share.

At the $n$-profile $\overrightarrow{u}$ with one $u_{F}$ and $n-1$ others $%
u_{G}$ we have $\mathcal{W}(\overrightarrow{u})=1$ therefore inequality (\ref%
{13}) fails. Note that this failure is not a knife edge situation: the set
of profiles where the corresponding profile of MaxMin utilities is not
feasible is open in $%
\mathbb{R}
_{+}^{2^{A}}$.

For $ii)$ pick any partition $\pi $ and check the inequality $\Gamma
_{n}^{\pi }(u)=\frac{1}{n}(\overset{n}{u})(\pi )\geq MinMax_{n}(u)$ for all $%
u$. $\blacksquare \smallskip $

We note that both statements in Lemma 3 hold in the cake-cutting model with
very general preferences (\cite{BM1}) but there the proof of $ii)$ is much
harder!\smallskip

Our next result, technically very simple, shows an important benefit of
choosing a guarantee in the \textquotedblleft duality
interval\textquotedblright\ $[MinMax_{n}(u),MaxMin_{n}(u)]$.

Recall from section 3 (second paragraph) the notation $\mathcal{S}ub$ and $%
\mathcal{S}up$ for the sets of sub- and super-additive functions. For
instance in Example 1 Y's utility is subadditive, Z's is superadditive and
X's is neither.\smallskip

\textbf{Proposition 1 }\textit{Suppose the guarantee }$\Gamma _{n}\in 
\mathcal{G}(A;n)$\textit{\ is such that}%
\begin{equation}
\Gamma _{n}(u)\in \lbrack MinMax_{n}(u),MaxMin_{n}(u)]\text{ for all }u\in 
\mathcal{M}^{+}  \label{17}
\end{equation}

\noindent \textit{Then} $\Gamma _{n}(u)=\frac{1}{n}u(A)$\textit{\ if }$u\in 
\mathcal{A}dd$\textit{; }$\Gamma _{n}(u)\geq \frac{1}{n}u(A)$\textit{\ if }$%
u\in \mathcal{S}ub$\textit{; and }$\Gamma _{n}(u)\leq \frac{1}{n}u(A)$%
\textit{\ if }$u\in \mathcal{S}up$\textit{.\smallskip }

\textbf{Proof} If $u\in \mathcal{S}ub$ (resp. $u\in \mathcal{S}up$) we have $%
u(A)=\min_{\pi \in \mathcal{P}(n;A)}(\overset{n}{u})(\pi )$ (resp. $%
u(A)=\max_{\pi \in \mathcal{P}(n;A)}(\overset{n}{u})(\pi )$) hence $\frac{1}{%
n}u(A)=MinMax_{n}(u)$ (resp. $MaxMin_{n}(u)$) therefore (\ref{17}) implies
the desired inequalities. $\blacksquare \smallskip $

If $u$ is additive $MaxMin_{n}(u)=\frac{1}{n}u(A)$, so by statement $i)$ in
Lemma 3 the Proportional Share is the best possible guarantee and the
compelling interpretation of ex ante fairness.

Note that property (\ref{17}) is not very restrictive: it is clearly
satisfied by the Proportional Share, the fixed partitions guarantees $\Gamma
_{n}^{\pi }$, the D\&C and B\&S guarantees defined shortly, and their convex
combinations.

In subsection 11.1 of the Appendix we show that the guarantee of the naive
Multi Auction rule (auctioning objects one by one, see 5-th paragraph in
section 1) falls below the duality interval: it is often much smaller than
the $MinMax$ guarantee. We dismiss MA for this very reason.\smallskip

\textbf{Remark 3 }There is a precise connection between the duality interval
in (\ref{17}) and Envy Freeness, confirming the trade-off between ex ante
and ex post fairness in Remark 2 above. At an envy free allocation, it is
clear that every agent $i$\ gets at least her $MinMax_{n}(u_{i})$ utility.
Conversely\ if the single-valued rule $(\mathcal{M}^{+})^{N}\ni 
\overrightarrow{u}\rightarrow (\pi ,t)\in \mathcal{P}(N;A)\times \mathcal{T}%
(N)$\textit{\ }is efficient and envy-free, it must implement precisely the $%
MinMax$\ guarantee: we check that for each utility function $u_{i}$\ we can
complete a profile $(u_{i},u_{-i})$\ at which the rule gives to agent $i$\
precisely his $MinMax_{n}(u)$\ utility.

Fix $u_{1}\in \mathcal{M}^{+}$, $\pi =(S_{k})_{k=1}^{n}$ achieving $%
\min_{\pi \in \mathcal{P}(n;A)}[u_{1}](\pi )$, and a positive number $\delta 
$. Construct a profile where the common utility $v$ of the $n-1$ other
agents is the cover of the sequence $\{(S_{k},u_{1}(S_{k})+\alpha );k\in
\lbrack n]\}$. If $\alpha $ is very large any assignment of the shares $%
S_{k} $ to the agents is efficient (and any other efficient partition
distributes the same utilities pre-transfers). By the construction of
utility $v$, at an envy free and efficient allocation the transfers make
agent 1 indifferent between all the shares so her utility is $minMax_{n}(u)$.

\section{Two Divide\&Choose rules}

The two rules have the same guarantee and their building blocks are the $\pi 
$-auction and averaging-auction in section 4.\smallskip

\textbf{Definition 4 }\textit{Divide\&Choose}$_{n}^{\mathbf{1}}$

\noindent \textit{Stage 1: run a simple auction for the role of Divider; the
winner }$i^{\ast }$\textit{\ is (one of) the highest bidder(s) with a bid }$%
\beta _{i}\geq 0$\textit{;}

\noindent \textit{Stage 2:\ agent }$i^{\ast }$\textit{\ pays }$\frac{1}{n}%
\beta _{i}$\textit{\ to every other agent and picks a partition }$\pi ^{\ast
}=\{S_{k}\}_{k=1}^{n}$\textit{\ in }$\mathcal{P}(n,A)$;

\noindent \textit{Stage 3: run the }$\pi ^{\ast }$\textit{-auction between
all agents.}\smallskip

\textbf{Definition 5 }\textit{Divide\&Choose}$_{n}^{\mathbf{2}}$

\noindent \textit{Stage 1:\ each agent }$i$\textit{\ picks a partition }$\pi
^{i}$ \textit{in} $\mathcal{P}(n,A)$\textit{;}

\noindent \textit{Stage 2: run the averaging auction between the guarantees }%
$\Gamma _{n}^{\pi ^{i}},i\in N$.\smallskip

The D\&C$_{n}^{2}$ rule takes longer to run than D\&C$_{n}^{1}$ because the
averaging auction will first identify a partition $\pi ^{\widehat{i}}$
maximising $\sum_{i\in N}\Gamma _{n}^{\pi ^{j}}(u_{i})$ over $j$ before
running the $\pi ^{\widehat{i}}$-auction.\smallskip

\textbf{Proposition 2 }

\noindent \textit{In the D\&C}$_{n}^{1}$\textit{\ rule, agent }$i$\textit{'s
play is safe \textbf{if and only if} he bids }$\beta
_{i}=MaxMin_{n}(u_{i})-MinMax_{n}(u_{i})$\textit{\ in stage 1; chooses if he
wins a partition }$\pi ^{\ast }$\textit{\ maximising }$(\overset{n}{u_{i}}%
)(\pi )$\textit{\ in stage 2, and reports truthful equalising transfers
across the shares of }$\pi ^{\ast }$\textit{\ in stage 3.}

\noindent \textit{In the D\&C}$_{n}^{2}$\textit{\ rule, agent }$i$\textit{'s
play is safe \textbf{if and only if }she proposes a partition }$\pi ^{i}$%
\textit{\ maximising }$(\overset{n}{u_{i}})(\pi )$\textit{\ in stage 1, then
reports truthful equalising transfers across the guarantees }$\Gamma
_{n}^{\pi ^{j}}(u_{i}),j\in N$\textit{, and finally reports truthful
transfers in the final }$\pi ^{\widehat{i}}$\textit{-auction.}

\noindent \textit{Both rules implement the guarantee}%
\begin{equation*}
\Gamma _{n}^{DC}(u)=\frac{1}{n}MaxMin_{n}(u)+\frac{n-1}{n}MinMax_{n}(u)
\end{equation*}%
\begin{equation}
=\frac{1}{n^{2}}\max_{\pi \in \mathcal{P}(n,A)}u(\pi )+\frac{n-1}{n^{2}}%
\min_{\pi \in \mathcal{P}(n,A)}u(\pi )  \label{37}
\end{equation}

\noindent \textit{The guarantees }$\Gamma _{n}^{DC}$\textit{\ is Positive,
Responsive, and in the duality interval (\ref{17}).\smallskip }

\textbf{Proof }\textit{For D\&C}$_{n}^{1}$\textit{.} In the $\pi $-auction
agent $i$ guarantees the utility $\frac{1}{n}(\overset{n}{u})(\pi )$ (Lemma
1). So as the Divider her best choice of $\pi $ guarantees the utility $%
\max_{\pi }\frac{1}{n}(\overset{n}{u})(\pi )=MaxMin_{n}(u_{i})$ (Definition
3). As a Chooser, the worst possible choice of $\pi $ by the Divider gives $%
\min_{\pi }\frac{1}{n}(\overset{n}{u})(\pi )=MinMax_{n}(u_{i})$ to our
agent. So the worst drop in guaranteed utility between the roles of Divider
and Chooser is $\delta _{i}=MaxMin_{n}(u_{i})-MinMax_{n}(u_{i})$.

Her bid $x_{i}$ in stage 1 secures the utility $MaxMin_{n}(u_{i})-\frac{n-1}{%
n}x_{i}$ if it wins, and $MinMax_{n}(u_{i})+\frac{1}{n}x_{i}$ if it loses:
bidding $\delta _{i}$ maximises the smallest of these two, and her final
guarantee is as announced in (\ref{37}).

\textit{For D\&C}$_{n}^{2}$. Agent $i$'s guaranteed utility in stage 2 is $%
\frac{1}{n}\sum_{j\in \lbrack n]}\frac{1}{n}(\overset{n}{u_{i}})(\pi ^{j})$
(Lemma 2 ) so the worst case is when $(\overset{n}{u_{i}})(\pi
^{j})=\min_{\pi \in \mathcal{P}(n;A)}(\overset{n}{u_{i}})(\pi )$ for each $%
j\neq i$. Therefore proposing in stage 1 an optimal partition $\pi ^{i}$
securing $\max_{\pi }\frac{1}{n}(\overset{n}{u})(\pi )$ delivers the same
guarantee (\ref{37}).

We omit the easy proof that no other play is safe in either \ version of D\&C%
$_{n}$. $\blacksquare $\smallskip

\textbf{Example 1 }\textit{(continued)} \textbf{for D\&C}$_{3}^{1}$

From the $MaxMin$ and $MinMax$ values in (\ref{38}) we have%
\begin{equation*}
\begin{array}{ccc}
\Gamma _{3}^{DC}(u_{X}) & \Gamma _{3}^{DC}(u_{Y}) & \Gamma _{3}^{DC}(u_{Z})
\\ 
6 & 7\frac{2}{3} & 3\frac{2}{3}%
\end{array}%
\end{equation*}

We compute the allocation reached by the safe play of all three agents.

In D\&C$^{1}$ the bids in stage 1 are $(3,5,5)$ for X,Y and Z respectively.
The way we break ties between Y and Z is now critical. If Z is chosen as the
Divider, he pays $1\frac{2}{3}$ to X and to Y, then picks the bundle
partition $\pi ^{PS}$ where his safe bid of $21$ wins and he gives an extra $%
7$ to X and to Y. Final allocation and utilities are%
\begin{equation*}
\begin{array}{ccc}
\text{X} & \text{Y} & \text{Z} \\ 
(\varnothing ,8\frac{2}{3}) & (\varnothing ,8\frac{2}{3}) & (A,-17\frac{1}{3}%
) \\ 
8\frac{2}{3} & 8\frac{2}{3} & 3\frac{2}{3}%
\end{array}%
\end{equation*}%
where Z gets nothing more than his guaranteed utility.

If instead Y wins stage 1, in stage 2 she pays $1\frac{2}{3}$ to X and to Z
then can safely divide $A$ either as $\pi ^{\ast }=\{ac,b,\varnothing \}$ or 
$\pi ^{\ast \ast }=\{bc,a,\varnothing \}$. Say she chooses $\pi ^{\ast }$.
We computed in subsection 4.1 the corresponding (safe and truthful) transfer
reports (\ref{41}) and the resulting allocation (\ref{42}). To the latter we
add the payments in stage 2. Final allocation and utilities%
\begin{equation*}
\begin{array}{ccc}
\text{X} & \text{Y} & \text{Z} \\ 
(ac,-2) & (b,-5) & (\varnothing ,7) \\ 
10 & 10 & 7%
\end{array}%
\end{equation*}%
a serious Pareto improvement over the choice of Z as winner in stage 1,
reflecting the fact that $\pi ^{\ast }$ delivers $6$ more units of total
surplus than $\pi ^{PS}$.

If agent Y after winning stage 1 chooses instead the efficient (unbeknownst
to her) partition $\pi ^{\ast \ast }$ we let the reader check the final
result

\begin{equation*}
\begin{array}{ccc}
\text{X} & \text{Y} & \text{Z} \\ 
(bc,-3) & (a,-5) & (\varnothing ,8) \\ 
12 & 10 & 8%
\end{array}%
\end{equation*}%
yet another Pareto improvement over the previous choice of Y.\smallskip

\textbf{Example 1}\textit{\ (continued)}\textbf{\ for D\&C}$_{3}^{2}$

To agent Z the partition with the best guarantee is the bundle $\pi ^{PS}$.
For X the best choice is $\pi ^{\ast \ast }=\{bc,a,\varnothing \}$ with a
guarantee of $8$, but Y has a choice between $\pi ^{\ast
}=\{ac,b,\varnothing \}$ and $\pi ^{\ast \ast }$.

Assuming that Y picks $\pi ^{\ast }$ in stage 1, we compute the $3\times 3$
matrix of guarantees $\Gamma _{3}^{\pi }(u_{i})=\frac{1}{3}(\overset{3}{u_{i}%
})(\pi )$ and the corresponding safe balanced transfers:%
\begin{equation*}
\lbrack \frac{1}{3}(\overset{3}{u_{i}})(\pi )]:%
\begin{array}{cccc}
& \pi ^{\ast \ast } & \pi ^{\ast } & \pi ^{PS} \\ 
\text{X} & 8 & 6 & 5 \\ 
\text{Y} & 11 & 11 & 6 \\ 
\text{Z} & 4 & 3 & 7%
\end{array}%
\Longrightarrow \lbrack t_{i}^{\pi }]:%
\begin{array}{cccc}
& \pi ^{\ast \ast } & \pi ^{\ast } & \pi ^{PS} \\ 
\text{X} & -1\frac{2}{3} & +\frac{1}{3} & +1\frac{1}{3} \\ 
\text{Y} & -1\frac{2}{3} & -1\frac{2}{3} & +3\frac{1}{3} \\ 
\text{Z} & +\frac{2}{3} & +1\frac{2}{3} & -2\frac{1}{3}%
\end{array}%
\end{equation*}

The surplus maximising partition is $\pi ^{\ast \ast }$: the sum of its
column in the right (resp. left) matrix is minimal at $-2\frac{2}{3}$ (resp.
maximal at $23$). So before running the $\pi ^{\ast \ast }$-auction, we
perform transfers $\tau $ as in the $\pi ^{\ast \ast }$ column, augmented by
a share $\frac{1}{3}|2\frac{2}{3}|$ of the slack for each agent: $\tau =(-%
\frac{7}{9},-\frac{7}{9},+1\frac{5}{9})$. Then the $\pi ^{\ast \ast }$%
-auction delivers the allocation X: $(bc,-4\frac{2}{3})$; Y: $(a,-1\frac{2}{3%
})$; Z: $(\varnothing ,6\frac{1}{3})$ which we finally combine with $\tau $:%
\begin{equation*}
\begin{array}{ccc}
\text{X} & \text{Y} & \text{Z} \\ 
(bc,-5\frac{4}{9}) & (a,-2\frac{4}{9}) & (\varnothing ,7\frac{8}{9}) \\ 
9\frac{5}{9} & 12\frac{5}{9} & 7\frac{8}{9}%
\end{array}%
\end{equation*}

We conclude that safe reporting in the two versions of D\&C delivers
significantly different allocations, all the more so if some agents have
several choices of optimal partitions.\smallskip

Finally we comment on an unappealing feature of D\&C$^{1,2}$. In the
reporting stages common to both rules each agent only reveals the relative
utilities between the shares of certain partitions but the level of his
absolute utility remains private: this increases privacy but is detrimental
to efficiency.

For instance if utility $u$\textit{\ }is additive the safe bid in D\&C$^{1}$%
\ is zero\ and any partition is a safe proposal in both rules (because $(%
\overset{n}{u})(\pi )=u(A)$\ for any $\pi $). Then if $u$ is so much higher
than other utilities that efficiency requires to give this agent all the
goods, her bid in D\&C$^{1}$ is still zero and some agent $2$ with non
additive utility will become the Divider; if $2$ does not bundle all goods
in one share, the final allocation is for sure inefficient.

On the contrary in the Bid\&Sell rule to which we now turn, individual
messages are related to the absolute utilities and avoid this type of
inefficiencies: this is formally proven by Proposition 5 in section 9.

\section{The Bid\&Sell rule}

For a non negative price vector $p\in 
\mathbb{R}
_{+}^{A}$ we use the same notation $p_{S}=\sum_{a\in S}p_{a}$ as if $p$
described an additive utiltity. We write $\Delta (x)$ for the simplex of
prices such that $p_{A}=x$. Because the recursive definitions of the B\&S
rule and its guarantee work over shrinking subsets of objects, we make
explicit their dependence on the set $A$.\smallskip

\textbf{Definition 6} \textit{Bid\&Sell for two agents}: B\&S$_{2}(A)$%
\newline
\textit{stage 1: each agent }$i$\textit{\ bids }$x_{i}$\textit{\ (a non
negative real number)} \textit{to become the Seller; (one of) the lowest
bidder(s) with bid }$x$\textit{\ becomes the Seller;}\newline
\textit{stage 2: the Seller chooses a price }$p\ $\textit{in }$\Delta (x)$;%
\newline
\textit{stage 3: the Buyer can buy any share }$S$\textit{\ of objects
(possibly }$\varnothing $\textit{\ or }$A$\textit{) at price }$p$; \textit{%
the Seller cashes the revenue and enjoys the unsold goods.}\newline
\textit{Final allocation: Buyer }$(S,-p_{S})$\textit{\ ; Seller }$%
(A\diagdown S,p_{S})$.\smallskip

To understand how to bid safely we compute first the safe utility $%
W_{2}(u;x|A)$ an agent with utility $u$ becoming the Seller after bidding $x$
can secure by choosing optimally the price offered to the Buyer and
expecting the worst purchase from that agent:%
\begin{equation}
W_{2}(u;x|A)=\max_{p\in \Delta (x)}\min_{\varnothing \subseteq T\subseteq
A}(u(T)+p_{A\diagdown T})=x+\max_{p\in \Delta (x)}\min_{\varnothing
\subseteq T\subseteq A}(u(T)-p_{T})  \label{28}
\end{equation}

We compare it with the safe utility $L_{2}(u;x|A)$ this agent can secure if
her bid $x$ loses \textit{by a hair }(to a bid just below $x$) so she
becomes the Buyer and is offered the worst possible price such that the
whole bundle $A$ costs $x$:%
\begin{equation}
L_{2}(u;x|A)=\min_{p\in \Delta (x)}\max_{\varnothing \subseteq S\subseteq
A}(u(S)-p_{S})  \label{29}
\end{equation}

Clearly $W_{2}(u;x|A)$ increases in $x$ while $L_{2}(u;x|A)$ decreases hence
the safe bid in stage 1 is $x^{\ast }$ such that $W_{2}(u;x^{\ast
}|A)=L_{2}(u;x^{\ast }|A)$, which we show below is well defined. This common
value is the Bid \& Sell guarantee $\Gamma _{2}^{BS}(u|A)$.\smallskip

Even with three goods and two agents the computation of the bid functions $%
W_{2}$ and $L_{2}$ is a linear program harder to solve than computing the $%
MaxMin$ and $MinMax$ partitions as in section 5.\smallskip

In the next computation and in Example 2 after Definition 7 we use the
familiar notation%
\begin{equation*}
(z)_{+}=\max \{z,0\}\text{ ; }(z)_{-}=\min \{z,0\}
\end{equation*}

\textbf{Example 1 }\textit{(continued)}

For agent X involved\ in a \textbf{two person} division of $A$ the
guaranteed utility after a winning bid $x$ is%
\begin{equation*}
W_{2}(u_{X};x|A)=x+\max_{p\in \Delta (x)}\min
\{0,9-p_{a},6-p_{b},-p_{c},15-p_{ab},12-p_{ac},15-p_{bc},15-x\}
\end{equation*}

We can drop the two dominated terms $15-p_{ab}$ and $15-p_{bc}$, then check
that for $p=(\frac{3}{5}x,\frac{2}{5}x,0)$ the $\max \min $ term is $%
(15-x)_{-}$ and that this price is optimal. Therefore $W_{2}(u_{X};x|A)=\min
\{x,15\}$.

Next we compute $L_{2}(u_{X};x|A)$, the guaranteed utility after a losing
bid $x$:%
\begin{equation*}
L_{2}(u_{X};x|A)=\min_{p\in \Delta (x)}\max
\{0,9-p_{a},6-p_{b},-p_{c},15-p_{ab},12-p_{ac},15-p_{bc},15-x\}
\end{equation*}%
where we can only drop the term $15-x$.

For $x\leq 3$ the price $p=(x,0,0)$ is optimal and $L_{2}(u_{X};x|A)=15-x$.
For $x\geq 3$ the optimal price solves $15-p_{ab}=12-p_{ac}=15-p_{bc}$ and $%
L_{2}(u_{X};x|A)=(14-\frac{2}{3}x)_{+}$. Finally the two functions intersect
at the safe bid $x^{\ast }=8\frac{2}{5}$, guaranteeing to agent X the
utility $u_{X}=8\frac{2}{5}$.

Similar computations, omitted for brevity, give for Y:%
\begin{equation*}
W_{2}(u_{Y};x|A)=\min \{x,18\}
\end{equation*}%
\begin{equation*}
L_{2}(u_{Y};x|A)=18-x\text{ on }[0,3]\text{ ; }=16\frac{1}{2}-\frac{1}{2}x%
\text{ on }[3,9]\text{ ; }=(18-\frac{2}{3}x)_{+}\text{ above }9
\end{equation*}%
and these two functions intersect at the safe bid $x^{\ast }=11\frac{1}{4}$
guaranteeing the utility $11\frac{1}{4}$.

For agent Z we find similarly%
\begin{equation*}
W_{2}(u_{Z};x|A)=x\text{ on }[0,6]\text{ ; }=\frac{1}{2}x+3\text{ on }[6,18]%
\text{ ;}
\end{equation*}%
\begin{equation*}
=\frac{1}{3}x+6\text{ on }[18,45]\text{ ; }=21\text{ above }45
\end{equation*}%
\begin{equation*}
L_{2}(u_{Z};x|A)=(21-x)_{+}
\end{equation*}%
so that Z's safe bid is $x^{\ast }=12$ for the guaranteed utility $9$.

We find that the B\&S guarantees improve those of the D\&C$_{2}$ rule in a%
\textit{\ }\textbf{two person} division of $A$ (Proposition 2) for all three
agents 
\begin{equation*}
\begin{array}{ccc}
& \Gamma _{2}^{DC} & \Gamma _{2}^{BS} \\ 
\text{X} & 6\frac{1}{2} & 8\frac{2}{5} \\ 
\text{Y} & 8\frac{1}{2} & 11\frac{1}{4} \\ 
\text{Z} & 4\frac{1}{2} & 9%
\end{array}%
\end{equation*}

This pattern is of course not a general feature of the comparison between
D\&C and B\&S.\smallskip

For a larger number $n$ of agents, the rule $B\&S_{n}(A)$ is defined
recursively, through at most $n-1$ rounds of bidding: in each round one
agent is the Buyer and the remaining other agents are Sellers; the Buyer
leaves after buying some goods (perhaps none) from all the Sellers.
Naturally the computational difficulty increases sharply.\smallskip

\textbf{Definition 7 }B\&S$_{n}(A)$: Bid\&Sell for $n\geq 3$\newline
Suppose the rule B\&S$_{\upsilon }(B)$ is already defined for $|B|\leq n-1$
and define B\&S$_{n}(A)$ as follows.

\textit{Stage 1: each agent }$i$\textit{\ bids }$x_{i}$\textit{\ to become
Seller or Buyer; (one of) the highest bidder(s) becomes the Buyer;}\newline
\textit{Stage 2: each of the }$n-1$\textit{\ Sellers }$j$\textit{\ chooses a
price }$p_{j}$\textit{\ in }$\Delta (x_{j})$;\newline
\textit{Stage 3: the Buyer buys a share }$S$\textit{\ of goods by paying }$%
p_{j}(S)$\textit{\ to \textbf{each} Seller and leaves; the rule stops if }$%
S=A$\textit{, otherwise we go to}\newline
\textit{Stage 4: the remaining agents play }B\&S$_{n-1}(A\diagdown S)$.$%
\smallskip $

The worst utility $W_{n}(u;x|A)$ from becoming a Seller after bidding $x$ is
now%
\begin{equation}
W_{n}(u;x|A)=\max_{p\in \Delta (x)}\min_{\varnothing \subseteq T\subseteq
A}(\Gamma _{n-1}^{BS}(u|T)+p_{A\diagdown T})=x+\max_{p\in \Delta
(x)}\min_{\varnothing \subseteq T\subseteq A}(\Gamma _{n-1}^{BS}(u|T)-p_{T})
\label{30}
\end{equation}%
and the worst utility as a Buyer after bidding $x$ is%
\begin{equation}
L_{n}(u;x|A)=\min_{p\in \Delta ((n-1)x)}\max_{\varnothing \subseteq
S\subseteq A}(u(S)-p_{S})=\min_{p\in \Delta (x)}\max_{\varnothing \subseteq
S\subseteq A}(u(S)-(n-1)p_{S})  \label{31}
\end{equation}%
because the worst case is when the $n-1$ other bids are just below $x$%
.\smallskip

\textbf{Lemma 4 }\textit{For any non null utility }$u\in \mathcal{M}^{+}$%
\textit{\ the recursive programs (\ref{30}),(\ref{31}), together with the
initial pair (\ref{28}), (\ref{29}), define unambiguously}\newline
\textit{the function }$W_{n}(u;x|A)$\textit{\ concave and strictly
increasing in }$x$\textit{\ from }$0$\textit{\ to }$u(A)$;\newline
\textit{the function }$L_{n}(u;x|A)$\textit{\ convex and strictly decreasing
in }$x$\textit{\ from }$u(A)$\textit{\ to }$0$;\newline
\textit{and the guarantee }$\Gamma _{n}^{BS}(u)$\textit{\ at their
intersection: }$W_{n}(u;x^{\ast }|A)=L_{n}(u;x^{\ast }|A)=\Gamma
_{n}^{BS}(u|A)$.$\mathit{\smallskip }$

These properties imply: $0<\Gamma _{n}^{BS}(u|A)<u(A)$. In particular the
buyer in stage 3 buys at least one good.\smallskip

\textbf{Proposition 3 }\textit{The guarantee }$\Gamma _{n}^{BS}$\textit{\ is
Positive, Responsive, and in the duality interval (}\ref{17}).$\smallskip $

The proof of the key Lemma 4 and its corollary Proposition 3, in subsection
11.2 of the Appendix, is a non trivial application of the minimax
theorem.\smallskip

We illustrate the recursion defining the B\&S$_{3}$ rule in a simpler
instance than Example 1.\smallskip

\textbf{Example 2 }\textit{Three agents F, H, K share three identical goods
and their utilities are}%
\begin{equation*}
\begin{array}{cccc}
\text{\# of goods} & \text{1} & \text{2} & \text{3} \\ 
\text{F} & 5 & 5 & 5 \\ 
\text{H} & 0 & 4 & 6 \\ 
\text{K} & 1 & 3 & 6%
\end{array}%
\end{equation*}

So F\ is a Frugal agent who needs not more than one good, K is superadditive
and H is neither sub- nor super-additive.

As the goods are identical, we use the fact that the optimal price $p$ in (%
\ref{30}), (\ref{31}) can be taken symmetric over the goods (Lemma 10 in
section 11.3).

Before computing the two functions $W_{3}$, $L_{3}$ for a utility $u$ we
must retrieve the two person guarantees $\Gamma _{2}^{BS}(u|k)$ when only $k$
goods are available, $k=1,2,3$. We computed this for the Frugal agent in
Example 0 section 1: after scaling up 5 times those earlier results we have%
\begin{equation*}
\Gamma _{2}^{BS}(F|3)=3\frac{3}{4}\text{ ; }\Gamma _{2}^{BS}(F|2)=3\frac{1}{3%
}\text{ ; }\Gamma _{2}^{BS}(F|1)=1\frac{2}{3}
\end{equation*}%
then we can apply (\ref{30}), (\ref{31}):%
\begin{equation*}
W_{3}(F;x|3)=\min \{x,\frac{5}{3}+\frac{2}{3}x,\frac{10}{3}+\frac{1}{3}x,%
\frac{15}{4}\}=\min \{x,3\frac{3}{4}\}
\end{equation*}%
\begin{equation*}
L_{3}(F;x|3)=\max \{0,5-\frac{2}{3}x,5-\frac{4}{3}x,5-2x\}=(5-\frac{2}{3}%
x)_{+}
\end{equation*}

Agent F's safe bid in Stage 1 of B\&S$_{3}$, at the intersection of these
two functions, is $x_{F}^{\ast }=3$. Her guaranteed surplus is also $\Gamma
_{3}^{BS}(F)=3$.

The same computations for agent H start with the two person problems with 1,
2 or 3 goods. For instance the two functions%
\begin{equation*}
W_{2}(H;x|3)=\min \{x,\frac{2}{3}x,4+\frac{1}{3}x,6\}=\min \{\frac{2}{3}x,6\}
\end{equation*}%
\begin{equation*}
L_{2}(H;x|3)=\max \{0,4-\frac{2}{3}x,6-x\}=(6-x)_{+}
\end{equation*}%
intersect at $x=\frac{18}{5}$ and $\Gamma _{2}^{BS}(H|3)=\frac{12}{5}$. We
find similarly $\Gamma _{2}^{BS}(H|2)=\frac{4}{3}$, $\Gamma _{2}^{BS}(H|1)=0$%
. Then we compute%
\begin{equation*}
W_{3}(H;x|3)=\min \{\frac{2}{3}x,\frac{4}{3}+\frac{1}{3}x,\frac{12}{5}%
\}=\min \{\frac{2}{3}x,\frac{12}{5}\}
\end{equation*}%
\begin{equation*}
L_{3}(H;x|3)=\max \{0,4-\frac{4}{3}x,6-2x\}=(6-2x)_{+}
\end{equation*}%
and conclude that H's safe bid in stage 1 of B\&S$_{3}$ is $x_{H}^{\ast }=2%
\frac{1}{4}$ guaranteeing $\Gamma _{3}^{BS}(H)=1\frac{1}{2}$.

Agent K two person guarantees are computed as $\Gamma _{2}^{BS}(K|3)=3$, $%
\Gamma _{2}^{BS}(K|2)=1\frac{1}{2}$, $\Gamma _{2}^{BS}(K|1)=\frac{1}{2}$,
and her safe bid is $x_{K}^{\ast }=2\frac{1}{16}$ guaranteeing $\Gamma
_{3}^{BS}(K)=1\frac{7}{8}$.

The largest bid in stage 1 is $x_{F^{5}}^{\ast }=3$ so F is the first buyer.
In stage 2 agents H,K choose equal unit prices for the 3 goods, respectively 
$p^{H}=\frac{1}{3}x_{H}^{\ast }=\frac{3}{4}$ and $p^{K}=\frac{1}{3}%
x_{K}^{\ast }=\frac{11}{16}$. In stage 3 agent F pays $1\frac{7}{16}$ for
one good and her final utility is $5-1\frac{7}{16}=3\frac{9}{16}$.

In stage 4 agents H and K play B\&S$_{2}$ for the two remaining goods. Agent
H bids $2\frac{2}{3}$, larger than K's bid $\frac{1}{2}$ so H is the next
buyer: he buys both goods and pays $1\frac{1}{2}$ to K. The final allocation
is efficient: one good to F and two to H, for the final utilities%
\begin{equation*}
F:u=3\frac{9}{16}\text{ ; }H:u=3\frac{1}{4}\text{ ; }K:u=2\frac{3}{16}
\end{equation*}%
where F' and K's share of surplus are less than 20\% larger than their
respective guaranteed shares (respectively $\Gamma _{3}^{BS}(F)=3$ and $%
\Gamma _{3}^{BS}(K)=1\frac{30}{16}$) whereas H more than doubles his
guarantee $\Gamma _{3}^{BS}(H)=1\frac{1}{2}$.

\section{Comparing the B\&S, D\&C, and PS guarantees}

\subsection{More common properties}

We already know that all three guarantees are Positive and in the duality
interval, and\ that $\Gamma ^{BS}$ and $\Gamma ^{DC}$ (the same guarantee
for both D\&C rules) are Responsive.\smallskip

\textbf{Lemma 5}

\noindent $i)$ \textit{The guarantees }$\Gamma _{n}^{BS},\Gamma _{n}^{DC}$%
\textit{\ and }$\Gamma _{n}^{PS}$\textit{\ are continuous and weakly
increasing in the individual utility }$u$\textit{.}

\noindent $ii)$\textit{\ They are also scale invariant (}$\Gamma
_{n}(\lambda u)=\lambda \Gamma _{n}(u)$\textit{\ for }$\lambda >0$\textit{),
weakly increasing in }$A$\textit{\ and weakly decreasing in }$n$\textit{.
For all }$A,n,u$\textit{\ we have}%
\begin{equation*}
\Gamma _{n}(u|A)\leq \Gamma _{n}(u|A\cup a)\text{ and }\Gamma
_{n+1}(u|A)\leq \Gamma _{n}(u|A)
\end{equation*}

\textbf{Proof }Statement $i)$ is clear for $\Gamma _{n}^{DC}$ and $\Gamma
_{n}^{PS}$. For $\Gamma _{n}^{BS}$ both functions $W_{n}(u;\cdot )$ and $%
L_{n}(u;\cdot )$ increase weakly in $u$, so their intersection does too.

\textit{Statement }$ii)$\textit{\ for }$\Gamma _{n}^{DC}$

\noindent Scale invariance is clear. For the monotonicity in $A$ one checks
easily that both $MaxMin_{n}(u|A)$ and $MinMax_{n}(u|A)$ increase weakly in $%
A$. For the monotonicity in $n$ we fix $\pi ^{\ast }\in \mathcal{P}(n+1;A)$
and pick a share $S$ in $\pi ^{\ast }$ such that $u(S)\leq \frac{1}{n+1}(%
\overset{n+1}{u})(\pi ^{\ast }|A)$ (e. g. an empty share, if any). This
implies $\frac{n}{n+1}(\overset{n+1}{u})(\pi ^{\ast }|A)\leq (\overset{n+1}{u%
})(\pi ^{\ast }|A)-u(S)\leq \max_{\mathcal{P}(n;A)}(\overset{n}{u})(\pi |A)$
and that $MaxMin_{n}(u)$ decreases weakly in $n$. Pick next $\widehat{\pi }%
\in \mathcal{P}(n;A)$ such that $(\overset{n}{u})(\widehat{\pi }|A)=\min_{%
\mathcal{P}(n;A)}(\overset{n}{u})(\pi |A)$ and note that $(\overset{n}{u})(%
\widehat{\pi }|A)=(\overset{n+1}{u})(\widetilde{\pi }|A)$ for the partition $%
\widetilde{\pi }$ adding an empty share to $\widehat{\pi }$, therefore%
\begin{equation*}
nMinMax_{n}(u|A)=(\overset{n}{u})(\widehat{\pi }|A)=(\overset{n+1}{u})(%
\widetilde{\pi }|A)\geq (n+1)MinMax_{n+1}(u|A)
\end{equation*}%
implying that $MinMax_{n}(u)$ is also weakly decreasing in $n$.

\textit{Statement }$ii)$\textit{\ for }$\Gamma _{n}^{BS}$

\noindent Checking Scale Invariance is routine. The monotonicity in $A$ is
proven in subsection 11.2 in the second paragraph of the proof of Lemma 4.

For the monotonicity in $n$: taking $T=A$ in the minimisation part of
program (\ref{30}) gives $W_{n}(u;x)\geq \Gamma _{n-1}^{BS}(u|A)$ for all $x$%
, and this holds in particular at the $x^{\ast }$ optimal in the problem
with $n$ agents. $\blacksquare \smallskip $

For an additive utility $u$\textbf{\ }the three\textbf{\ }rules share the
guarantee $\frac{1}{n}u(A)$ (Proposition 1). For B\&S$_{n}$, just like for BA%
$_{n}$, the only safe bid is $x^{\ast }(u)=\frac{1}{n}u(A)=\Gamma
_{n}^{BS}(u)$; and if this makes you the Seller the price $p_{a}=\frac{1}{n}%
u(a)$ is uniquely safe. The omitted proof checks by induction that $%
W_{n}(u;x)=\min \{x,u(A)\}$ and $L_{n}(u;x)=(u(A)-(n-1)x)_{+}$.

Recall from the discussion at the end of section 6 that, on the contrary, in
D\&C$_{n}^{1}$ the safe bid is zero so that any partition is a safe choice
for both versions of the rule.

\paragraph{Computational complexity}

The recursive computation of $\Gamma _{n}^{BS}(u|A)$ from $\Gamma
_{n-1}^{BS} $ solves the two LPs (\ref{30}), (\ref{31}) of size $2^{m}$. For
a general $n $ we solve a pair of LPs for each agent to go from $n-1$ to $n$
and this may happen in each of the $n-1$ steps of the full recursive
algorithm. Therefore the number of LPs grows as $n^{2}$ so the complexity
remains polynomial as long as the number of objects is fixed. We already
noticed that it is exponential in the number of goods unless the goods are
identical, as follows from the general result in \cite{NS}. The same
conclusions apply to either D\&C rule, where the only hard step is to
identify the partitions $\pi $ minimising or maximising the utilities $(%
\overset{n}{u})(\pi )$.

The easy case of identical goods is discussed in subsections 8.4 below and
11.3 in the Appendix.

\subsection{Divergence from the Proportional Share}

We turn to a different effect already illustrated in Examples 0 and 1: as
the utility function becomes more sudadditive or more superadditive, the B\&S%
$_{n}$ guarantee deviates more from the Proportional Share than the D\&C$%
_{n} $ does.\smallskip

\textbf{Proposition 4 }\textit{For all }$n$\textit{\ and all }$u\in \mathcal{%
M}^{+}$\textit{\ we have}%
\begin{equation*}
\frac{n}{(n-1)m+1}\leq \frac{\Gamma _{n}^{BS}(u)}{\frac{1}{n}u(A)}\leq \frac{%
n\times m}{n+m-1}
\end{equation*}%
\begin{equation*}
\frac{1}{n}\leq \frac{\Gamma _{n}^{DC}(u)}{\frac{1}{n}u(A)}\leq \frac{\min
\{m,n\}+n-1}{n}
\end{equation*}%
\textit{In both cases the bounds are achieved at }$u_{G}$\textit{\ and }$%
u_{F}$\textit{\ respectively.\smallskip }

We see that the upper bound of $\frac{\Gamma _{n}^{BS}}{\Gamma _{n}^{PS}}$
is strictly larger than that of $\frac{\Gamma _{n}^{DC}}{\Gamma _{n}^{PS}}$,
with a single exception at $n=m=2$. And the lower bound of $\frac{\Gamma
_{n}^{BS}}{\Gamma _{n}^{PS}}$ is strictly lower than that of $\frac{\Gamma
_{n}^{DC}}{\Gamma _{n}^{PS}}$ if $m\geq n+2$, strictly larger if $m\leq n$,
and equal if $m=n+1$.

Moreover the ratio $\frac{\Gamma _{n}^{DC}}{\Gamma _{n}^{PS}}$ is always
below $2$, while $\frac{\Gamma _{n}^{BS}}{\Gamma _{n}^{PS}}$ can be
arbitrarily large.\smallskip

\textbf{Proof }We apply Lemma 5 twice. Every utility $u$ in $\mathcal{M}^{+}$
s. t. $u(A)=1$ satisfies $u_{G}\leq u\leq u_{F}$ and $\Gamma
_{n}^{BS},\Gamma _{n}^{DC}$ increase weakly in $u$, therefore%
\begin{equation*}
\Gamma _{n}^{M}(u_{G})\leq \Gamma _{n}^{M}(u)\leq \Gamma _{n}^{M}(u_{F})%
\text{ where }M\text{ is }D\&C\text{ or }B\&S
\end{equation*}

By scale invariance it is enough to show that $u_{G}$\textit{\ and }$u_{F}$
achieve the announced bounds for the two rules. If $M=D\&C$ it follows by
Proposition 2 after checking $\max_{\pi }(\overset{n}{u_{G}})(\pi )=1$, $%
\min_{\pi }(\overset{n}{u_{G}})(\pi )=0$ and%
\begin{equation*}
\max_{\pi }(\overset{n}{u_{F}})(\pi )=\min \{m,n\}\text{ ; }\min_{\pi }(%
\overset{n}{u_{F}})(\pi )=1
\end{equation*}

For $M=B\&C$ we use the more general result about dichotomous utilities in
Lemma 7 two subsections below. $\blacksquare $

\subsection{A revealing example}

Here the D\&C guarantee is unpalatable because it ignores important aspects
of the externalities across objects. This critique is more subtle than --but
similar to --that of the Proportional Share by the way it treats Greedy and
Frugal.\smallskip

\textbf{Example 3} Two agents, Abstemious and Choosy, share $4\ell $ goods
partitioned as four subsets, each with $\ell $ objects: $A=R\cup R^{\ast
}\cup L\cup L^{\ast }$. Think of two types of right gloves and two types of
left gloves.

Abstemious is happy with any pair of one right and one left glove: her
utility is the cover of $\{((r,\ell ),1)\}$ over the whole set $(R\cup
R^{\ast })\times (L\cup L^{\ast })$. Choosy wants no less than all gloves in 
$R\cup L$ or all in $R^{\ast }\cup L^{\ast }$: his utility is the cover of $%
\{(R\cup L,1)$, $(R^{\ast }\cup L^{\ast },1)\}$.

For both agents $MinMax_{2}(u)=0$, $MaxMin_{2}(u)=1$, so $\Gamma _{2}^{DC}$
gives $\frac{1}{2}$ to \textit{both} agents: the D\&C guarantee is
shockingly coarse, the more so as $\ell $ grows. By contrast we check that
the optimal bid in the B\&S$_{2}$ rule is $x^{\ast }=\frac{2\ell }{\ell +1}$
for \textit{both} agents (almost twice larger than $u(A)$). To Abstemious
this guarantees $\frac{\ell }{\ell +1}$ because her worst case as Seller is
to sell exactly $R\cup R^{\ast }$ or exactly $S\cup S^{\ast }$ for a net
utility $\frac{x}{2}$; and as Buyer there will be at least one pair costing
at most $\frac{2x}{m}$.

And to Choosy the B\&S$_{2}$ guarantee is $\frac{1}{\ell +1}$ because his
worst case as Seller is to sell exactly one glove in $R\cup R^{\ast }$ and
one in $S\cup S^{\ast }$ for a net utility $\frac{2x}{m}$; and as Buyer he
will have to pay $\frac{x}{2}$ to get any benefit.

\subsection{Identical goods}

Here the utility $u(S)$ depends only on the cardinality $s$ of the subset $S$
of goods: it is an increasing function $s\rightarrow u_{s}$ from $[m]$ into $%
\mathbb{R}
_{+}$. The \textit{median} of $u$, written $u_{med}$ is $u_{\frac{m}{2}}$ if 
$m$ is even, and $u_{med}=\frac{1}{2}(u_{\frac{m-1}{2}}+u_{\frac{m+1}{2}})$
if $m$ is odd.

In this rich class of utilities computing the D\&C$_{2}^{1}$ safe bid and
guarantee is fairly simple because $MaxMin_{2}(u)$ and $MinMax_{2}(u)$ are
respectively the maximum and minimum of $u_{s}+u_{m-s}$ over $s$. More work
is needed to compute them for the B\&S$_{2}$ rule without restrictions on
the sequence $(u_{s})_{[m]}$: the still simple programs are described in
Lemma 11, section 11.3 of the Appendix. Here we apply this result to
describe $\Gamma _{2}^{BS}$ for convex or concave utility functions, and
compare it to $\Gamma _{2}^{DC}$.\smallskip

\textbf{Lemma 6}

\noindent $i)$ \textit{Suppose }$u$\textit{\ is either convex or concave.
Then the optimal bid and guarantee in the D\&C}$_{2}$\textit{\ rule are}%
\begin{equation*}
x^{\ast }=|\frac{1}{2}u_{m}-u_{med}|\text{ ; }\Gamma _{2}^{DC}(u)=\frac{1}{2}%
u_{med}+\frac{1}{4}u_{m}
\end{equation*}

\noindent $ii)$ \textit{If }$u$\textit{\ is convex the optimal bid and
guarantee in the B\&S}$_{2}$\textit{\ rule are}%
\begin{equation*}
x^{\ast }=\max_{0\leq s\leq m}\{\frac{m}{m+s}(u_{m}-u_{m-s})\}\text{ ; }%
\Gamma _{2}^{BS}(u)=\frac{m}{m+s^{\ast }}u_{m-s^{\ast }}+\frac{s^{\ast }}{%
m+s^{\ast }}u_{m}
\end{equation*}

\noindent $iv)$ \textit{If }$u$\textit{\ is concave they are}%
\begin{equation*}
x^{\ast }=\max_{0\leq s\leq m}\frac{m}{m+s}u_{s}\text{ ; }\Gamma
_{2}^{BS}(u)=\frac{m}{m+s^{\ast }}u_{s^{\ast }}
\end{equation*}

Statement $i)$ is clear once we check that for a convex $u$: $MaxMin_{2}(u)=%
\frac{1}{2}u_{m}$ and $MinMax_{2}(u)=u_{med}$; and vice versa if $u$ is
concave.\smallskip

Finally we generalise Example 0 to the sequence of dichotomous utilities
connecting the Frugal and Greedy ones: we compute explicitly the D\&C and
B\&S guarantees and bids for arbitrary $n$ and $m$.

For each integer $\theta \in \lbrack m]$ the dichotomous utility $u^{\theta
} $ is satisfied with no less and no more than $\theta $ goods:%
\begin{equation*}
u^{\theta }(S)=1\text{ if }|S|\geq \theta \text{ ; }u^{\theta }(S)=0\text{
if }|S|<\theta
\end{equation*}%
Here $u^{1}=u_{F}$ and $u^{m}=u_{G}$.\smallskip

\textbf{Lemma 7 }\textit{For the dichotomous utilities above}

\noindent $i)$ \textit{The optimal bid and guarantee in the D\&C}$_{n}^{1}$%
\textit{\ rule are\footnote{%
We omit for easy reading the case $\frac{m}{n}<\theta <\frac{m}{n}+1$ where $%
^{\theta }x^{\ast }=\frac{1}{n}(\lfloor \frac{m}{\theta }\rfloor -1)$ and $%
\Gamma _{2}^{DC}(u^{\theta })=\frac{1}{n^{2}}(\lfloor \frac{m}{\theta }%
\rfloor +n-1)$.}}%
\begin{equation*}
^{\theta }x^{\ast }=1-\frac{1}{n}\text{ ; }\Gamma _{2}^{DC}(u^{\theta })=%
\frac{1}{n}(2-\frac{1}{n})\text{ if }\theta \leq \frac{m}{n}
\end{equation*}%
\begin{equation*}
^{\theta }x^{\ast }=\frac{1}{n}\lfloor \frac{m}{\theta }\rfloor \text{ ; }%
\Gamma _{2}^{DC}(u^{\theta })=\frac{1}{n^{2}}\lfloor \frac{m}{\theta }%
\rfloor \text{ if }\theta \geq \frac{m}{n}+1
\end{equation*}%
$ii)$ \textit{The optimal bid and guarantee in the B\&S}$_{n}$\textit{\ rule
are}%
\begin{equation*}
x^{\ast }=\frac{m}{m+1+(n-2)\theta }\text{ ; }\Gamma _{n}^{BS}(u^{\theta })=%
\frac{m+1-\theta }{m+1+(n-2)\theta }
\end{equation*}

For instance with five agents and twenty goods $\Gamma _{5}^{BS}$ dominates $%
\Gamma _{5}^{DC}$ for all values of $\theta $ except $18,19$ and $20$ when
they are both less than a quarter of the Proportional guarantee. The ratio $%
\frac{\Gamma _{5}^{BS}}{\Gamma _{5}^{PS}}$ decreases regularly from $4\frac{3%
}{20}$ while $\frac{\Gamma _{5}^{DC}}{\Gamma _{5}^{PS}}$ is never above $1%
\frac{4}{5}$.

The proof of statement $i)$ is routine once we compute%
\begin{equation*}
MaxMin_{5}(u^{\theta })=\min \{1,\frac{1}{n}\lfloor \frac{m}{\theta }\rfloor
\}
\end{equation*}%
\begin{equation*}
MinMax(u^{\theta })=\frac{1}{n}\text{ if }\theta <\frac{m}{n}+1\text{ ; }=0%
\text{ if }\theta \geq \frac{m}{n}+1
\end{equation*}

That of statement $ii)$ is in section 11.3 of the Appendix.

\section{Guaranteed collective welfare}

\subsection{Reducing the bargaining gap}

At any $n$-profile of utilities $\overrightarrow{u}$ the $n$-person
guarantee $\Gamma _{n}$ ensures a collective welfare not smaller than $%
\sum_{N}\Gamma _{n}(u_{i})$. We can evaluate the bite of our guarantee by
measuring the difference between the efficient surplus $\mathcal{W}(%
\overrightarrow{u})=\max_{\pi }\overrightarrow{u}(\pi )$ and that sum,
relative to the largest efficiency loss resulting from a misallocation of
the objects.

We call the interval $[\min_{\mathcal{P}(N;A)}\overrightarrow{u}(\pi ),\max_{%
\mathcal{P}(N;A)}\overrightarrow{u}(\pi )]$ the \textit{bargaining gap} of
the problem $(A,N,\overrightarrow{u})$ and we say that the guarantee $\Gamma
_{n}$ \textit{reduces the bargaining gap} if%
\begin{equation}
\min_{\pi \in \mathcal{P}(N;A)}\overrightarrow{u}(\pi )\leq \sum_{N}\Gamma
_{n}(u_{i})\leq \max_{\pi \in \mathcal{P}(N;A)}\overrightarrow{u}(\pi )\text{
for all }\overrightarrow{u}\in (\mathcal{M}^{+})^{N}  \label{33}
\end{equation}%
The right hand inequality is just the definition (\ref{13}) of a guarantee,
but the left hand inequality is not necessarily true.

Recall from Proposition 1 that $\Gamma ^{BS}$, $\Gamma ^{DC}$ and $\Gamma
^{PS}$ are not less than $MinMax_{n}(u_{i})$, therefore they guarantee the
collective welfare $\sum_{N}MinMax_{n}(u_{i})$. This lower bound is not
logically related to $\min_{\pi \in \mathcal{P}(N;A)}\overrightarrow{u}(\pi
) $.\footnote{%
If $A=\{a,b,c,d\}$, $u_{1}$ is the cover of $\{ab,bc,cd,ad\}$ and $u_{2}$ is
the cover of $\{ac,ad,bc,bd\}$ (all with value $1$), then $%
MinMax_{2}(u_{i})=0$ for $i=1,2$, but $\min_{\pi }\overrightarrow{u}(\pi )=1$
for all $\pi $. If $A=\{a,a^{\prime },b,b^{\prime }\}$, $u_{1}$ is the cover
of $\{a,a^{\prime }\}$ and $u_{2}$ is the cover of $\{b,b^{\prime }\}$, (all
with value $1$) then $\overrightarrow{u}(\pi )=0$ if each agent gets useless
goods, but $MinMax_{2}(u_{i})=\frac{1}{2}$ for $i=1,2$.} However, the PS
guarantee$\frac{1}{n}u(A)$ clearly meets (\ref{33}): by Proposition 1 again,
so do our guarantees $\Gamma ^{BS}$ and $\Gamma ^{DC}$ if the utilities are
subadditive.\smallskip

\textbf{Lemma 8}

\noindent $i)$ \textit{With two agents, }$n=2$\textit{, the Bid }\& \textit{%
Sell and Divide }\&\textit{\ Choose guarantees reduce the bargaining gap.}

\noindent $ii)$ \textit{With three or more agents, the Divide \& Choose
guarantee may not reduce the bargaining gap.\smallskip }

\textbf{Proof}

\textit{Statement }$i)$

\noindent \textit{Step 1: for B\&S}. Fix a profile $u,v$ where $u$'s optimal
bid $x^{\ast }$ wins against $v$'s larger optimal bid $y^{\ast }$. Let $p\in
\Delta (x^{\ast })$ be such that $\Gamma _{2}^{BS}(u)=L(u;x^{\ast
})=\max_{\varnothing \subseteq S\subseteq A}(u(S)-p_{S})$. We increase $p$
to some $q\in \Delta (y^{\ast })$ so that $\Gamma _{2}^{BS}(u)\geq
\max_{\varnothing \subseteq S\subseteq A}(u(S)-q_{S})$.

Also $\Gamma _{2}^{BS}(v)=W(v;y^{\ast })=\min_{\varnothing \subseteq
S\subseteq A}(v(S)-q_{S})+y^{\ast }=v(\overline{S})-q_{\overline{S}}+y^{\ast
}$ for some $\overline{S}$. Then $\Gamma _{2}^{BS}(u)\geq u(\overline{S}%
^{c})-q_{\overline{S}^{c}}$ so $\Gamma _{2}^{BS}(u)+\Gamma _{2}^{BS}(v)\geq
u(\overline{S}^{c})+v(\overline{S})$.\smallskip

\noindent \textit{Step 2: for D\&C}. Fix a profile $u,v$ and let $S,T$ be
such that $\min_{\pi }(\overset{n}{u})(\pi )=u(S)+u(S^{c})$ and $\min_{\pi
}[v](\pi )=v(T)+v(T^{c})$. The computation of $\Gamma _{2}^{DC}$
(Proposition 1) implies%
\begin{equation*}
\Gamma _{2}^{DC}(u)\geq \frac{1}{4}{\large (}u(T)+u(T^{c})+u(S)+u(S^{c})%
{\large )}
\end{equation*}%
and a similar lower bound for $\Gamma _{2}^{DC}(v)$. Summing up these
inequalities and rearranging gives the desired inequality (\ref{33}%
).\smallskip

\textit{Statement }$ii)$ Recall the dichotomous\textbf{\ }utilities $%
u^{\theta }$ and their guarantees in Lemma 7. A simple three person profile
violating (\ref{33}) for $\Gamma _{3}^{DC}$ has three goods and the profile $%
\overrightarrow{u}=(u^{2},u^{2},u^{1})$:%
\begin{equation*}
\Gamma _{3}^{DC}(u^{2})=\frac{1}{9}\text{ , }\Gamma _{3}^{DC}(u^{1})=\frac{5%
}{9}\text{ but }\min_{\pi \in \mathcal{P}(N;A)}\overrightarrow{u}(\pi )=1
\end{equation*}

$\blacksquare \smallskip $

We \textbf{conjecture} that the B\&S guarantee reduces the bargaining gap
for any $n$.

Our intuition comes again from the equation $\Gamma ^{BS}(u^{\theta })=\frac{%
m+1-\theta }{m+1+(n-2)\theta }$ in Lemma 7. At a profile $(u^{\theta
_{i}})_{N}$ the equality $\min_{\pi }\overrightarrow{u}(\pi )=1$ holds if
and only if $\sum_{i=1}^{n}\theta _{i}\leq m+n-1$. Then the left inequality
in (\ref{33}) follows from the convexity of $\theta \rightarrow \Gamma
^{BS}(u^{\theta })$ and is tight.

\subsection{When an agent's utility dominates}

For any two $u,v\in \mathcal{M}^{+}$ we say that $u$ \textit{dominates }$v$
(resp. dominates strictly) if we have%
\begin{equation*}
\max_{\varnothing \subseteq S\subseteq A}\partial _{a}v(S)\leq
\min_{\varnothing \subseteq S\subseteq A}\partial _{a}u(S)\text{ (resp. a
strict inequality) for all }a\in A
\end{equation*}

If in the profile $\overrightarrow{u}=(u_{i})_{i=1}^{n}$ utility $u_{1}$
dominates each $u_{i}$, $i\geq 2$, it is efficient to give all the goods to
agent $1$, strictly so if each domination is strict. This follows by
repeated application of the inequality $u_{1}(S)+u_{i}(T)\leq u_{1}(S\cup
a)+u_{i}(T\diagdown a)$ when $S,T$ are disjoint and $a\in T$.

Our last result reveals another serious advantage of the Bid \& Sell rule
over the Divide\&Choose rules.\smallskip

\textbf{Proposition 5 }\textit{Fix a profile }$\overrightarrow{u}%
=(u_{i})_{i=1}^{n}$\textit{\ where utility }$u_{1}$\textit{\ dominates
strictly }$u_{i}$\textit{\ for each }$i\geq 2$\textit{.}

\noindent $i)$ \textit{The B\&S}$_{n}$\textit{\ division rule where all
agents play safely implements the efficient outcome where agent }$1$\textit{%
\ eats all the goods.}

\noindent $ii)$\textit{\ The outcome of safe play in the D\&C rules may only
collect }$\frac{1}{n}$\textit{-th of the efficient surplus.\smallskip }

Proof of statement $i)$ in subsection 11.4.

For statement $ii)$ suppose for simplicity $m=n$, all goods are identical,
agent $1$ has the additive utility $u_{1}(S)=|S|$ and all utilities $u_{i}$
for $i\geq 2$ have marginals $\partial _{a}u_{i}(|S|)$ below $\varepsilon $,
with $\varepsilon <1$, and strictly decreasing in $|S|$. In D\&C$_{n}^{1}$
agent $1$ bids zero and the others bids are positive. The Divider will offer
the partition of $A$ in $n=m$ singletons and the surplus collected will be $%
1+(n-1)\varepsilon $, but the efficient surplus is $n$.

\section{Conclusion}

Our interpretation of ex ante fairness in terms of an individual guarantee
for each agent $u_{i}$ inside the benchmark interval $%
[MinMax_{n}(u_{i}),MaxMin_{n}(u_{i})]$ delivers the desired correction to
the coarse Proportional Share: a reward to a subadditive utility and a
penalty to a superadditive one (Proposition 1).

The Bid \& Sell and Divide \& Choose rules implement such guarantees
(Propositions 2 and 3), with B\&S responding more strongly than D\&C to sub-
and super-additivity (Proposition 4). It strongly outperforms D\&C when one
agent values each subset of goods much more than any other agent
(Proposition 5).

All agents playing safe in either rule does not in general extract the
efficient surplus because the messages reveal much less than full
utilities.Yet the price message in B\&S reveals more of the shape of the
Seller's utility than the partition in D\&C does of the Divider's. This
suggests that the safe play in B\&S captures more of the efficient surplus
than D\&C, at least on average.

Numerical experiments with two agents sharing up to seven goods in \cite{AK}
confirm this intuition. The B\&S rule captures on average between 95-99\% of
the efficient surplus whether both utilities are superadditive, both
subadditive, or one of each type. The corresponding range for the D\&S rule
(version 1 or 2) is 80-90\% for two subadditive agents, 65-75\% for two
superadditive ones, and 45-60\% for a mixed pair.

\section{Appendix: missing proofs}

\subsection{The Multi Auction rule}

Recall from section 1 that the MA simply runs $m$ independent Bundles
Auctions, one for each good. Each agent $i$ submits a profile of bids $\beta
^{i}\in 
\mathbb{R}
_{+}^{A}$; for each good $a$ (one of) the highest bidder(s) on $a$, agent $%
i^{\ast }$, gets $a$ and pays $\frac{1}{n}\beta _{i^{\ast }a}$ to every
other agent.

If utility $u_{i}$ is additive under MA the truthful bid $u_{ia}$ on each $a$
is the unique safe play and implement the PS guarantee. If all utilities are
additive the safe play by all picks an efficient allocation (that is even
Envy Free).

If the rule MA is used for general utilities in $\mathcal{M}^{+}$ the
marginal utility of adding $a$ to a subset of goods varies, so there is no
\textquotedblleft truthful\textquotedblright\ bid on $a$. For any utility $%
u\in \mathcal{M}^{+}$ the safe vector of bids solves the program:%
\begin{equation}
\Gamma _{n}^{MA}(u)=\max_{\beta \in 
\mathbb{R}
_{+}^{A}}{\large \{}\min_{\varnothing \subseteq S\subseteq A}(u(S)-\frac{n-1%
}{n}\beta _{S}+\frac{1}{n}\beta _{S^{c}}{\large \}}=\max_{\beta \in 
\mathbb{R}
_{+}^{A}}\min_{\varnothing \subseteq S\subseteq A}(u(S)-\beta _{S})+\frac{1}{%
n}\beta _{A}  \label{12}
\end{equation}

If our agent wins the auctions for the goods in $S$ and those only, she pays 
$\frac{n-1}{n}\beta _{a}$ for each $a$ in $S$, and gets in the worst case $%
\frac{1}{n}\beta _{a}$ for each $a$ outside $S$.

The guarantee $\Gamma _{n}^{MA}$ is neither Responsive nor Positive: $\Gamma
_{n}^{MA}(u_{G})=0<\frac{1}{n}=\Gamma _{n}^{MA}(u_{F})$. Indeed if Greedy's
bid $\beta $ is not zero, pick $a$ such that $\beta _{a}=\min_{b\in A}\beta
_{b}$, suppose Greedy wins all auctions except $a$ and check that his worst
utility is negative or zero. Next Frugal's safe bid of $\frac{1}{n}$ on
every good secures utility $\frac{1}{n}$ in the worst cases where she wins
all auctions or none of them.

Moreover $\Gamma _{n}^{MA}$ is dominated by the $MinMax$ guarantee, often
strictly so. To check the first claim pick $u\in \mathcal{M}^{+}$, a
partition $\pi =\{S_{k}\}_{k\in \lbrack n]}$ of $A$ and an optimal bid $%
\beta $ of $u$ in (\ref{12}). Then $\Gamma _{n}^{MA}(u)\leq (u(S_{k})-\beta
_{S_{k}})+\frac{1}{n}\beta _{A}$ for all $k$ and the sum of these
inequalities is $\Gamma _{n}^{MA}(u)\leq \frac{1}{n}(\overset{n}{u})(\pi )$.

An example where domination is strict is the utility $u=u_{F}+u_{G}$ when $%
m\geq 3$. One checks easily that $MaxMin_{2}(u)=MinMax_{2}(u)=1$ but $\Gamma
_{n}^{MA}(u)=\frac{m}{2(m-1)}$.

\subsection{Proof of Lemma 4 and Proposition 3}

Fixing $A$ and a single utility $u\in \mathcal{M}^{+}$, our first step is to
rewrite the programs (\ref{28}) (\ref{29}) in a more compact though less
transparent format using a well known combinatorial concept.

A vector $\delta =(\delta _{S})\in 
\mathbb{R}
_{+}^{2^{A\diagdown \varnothing }}$ is a \textit{balanced (set of) weights}
if for all $a\in A$ we have $\sum_{S:S\ni a}\delta _{S}=1$. We call $\delta $
\textit{minimal }if it is an extreme point of the convex compact set of
balanced weights, and write $\mathcal{B}_{m}$ the set of minimal balanced
weights for $m$ goods.\footnote{%
The size of $\mathcal{B}$ grows astronomically fast with $m$: $|\mathcal{B}%
|=2$ for $m=2$, $=6$ for $m=3$, $=27$ for $m=4$ and more than $15,000$ for $%
m=5$: see \cite{Gra}.} The simplest elements of $\mathcal{B}_{m}$ come from
the \textit{true partitions} $\{S_{k}\}$ of $A$: those where each $S_{k}$ is
non empty, $\delta _{S_{k}}=1$ for each $k$, and all other weights are $0$.
Let $\mathcal{B}_{m}^{\ast }$ be $\mathcal{B}_{m}$ minus the balanced
weights coming from the trivial partition $\{A\}$ ($\delta _{A}=1$ and other 
$\delta _{S}=0$).

Write the total weight of $\delta $ as $\overline{\delta }=\sum_{\varnothing
\neq S\subseteq A}\delta _{S}$. Then $\overline{\delta }>1$ for each $\delta 
$ in $\mathcal{B}_{m}^{\ast }$. The smallest of these sums is $\overline{%
\delta }=\frac{m}{m-1}$ when $\delta _{A\diagdown a}=\frac{1}{m-1}$ for all $%
a$ (all other weights are zero), and the largest one is $\overline{\delta }%
=m $ when $\delta _{a}=1$ for all $a$. Both claims follow from the identity $%
\sum_{S\varsubsetneq A}|S|\times \delta _{S}=m$.\smallskip

\textbf{Lemma 9}

\noindent The programs (\ref{28}) (\ref{29}) can be rewritten as follows:%
\begin{equation}
W_{2}(u;x)=\min \{x,u(A),\min_{\mathcal{B}_{m}^{\ast }}\frac{1}{\overline{%
\delta }}(\delta \cdot u-x)+x\}  \label{32}
\end{equation}%
\begin{equation}
L_{2}(u;x)=\max \{0,u(A)-x,\max_{\mathcal{B}_{m}^{\ast }}\frac{1}{\overline{%
\delta }}(\delta \cdot u-x)\}  \label{34}
\end{equation}

\textbf{Proof}.

\noindent We write $\nabla (Z)$ for the set of convex weights on $Z$, and
first rewrite the MaxMin expression in (\ref{28}):%
\begin{equation*}
W_{2}(u;x)-x=\max_{p\in \Delta (x)}\min_{\varnothing \subseteq T\subseteq
A}(u(T)-p_{T})=\max_{p\in \Delta (x)}\min_{\xi \in \nabla (2^{A})}\sum_{T\in
2^{A}}\xi _{T}(u(T)-p_{T})
\end{equation*}%
where $\xi $ has two coordinates $\xi _{A}$ and $\xi _{\varnothing }$.

Note that the mapping $\nabla (2^{A})\ni \xi \rightarrow \zeta \in 
\mathbb{R}
^{A}:\zeta _{a}=\sum_{T:a\in T}\xi _{T}$ is onto $[0,1]^{A}$, and apply the
minimax theorem to rewrite the last maxmin term above as%
\begin{equation}
\min_{\xi \in \nabla (2^{A})}\max_{a\in A}\sum_{T\in 2^{A}}\xi
_{T}u(T)-x\zeta _{a}\Longrightarrow W_{2}(u;x)=\min_{\xi \in \nabla
(2^{A})}\sum_{T\in 2^{A}}\xi _{T}u(T)+x(1-\min_{a}\varsigma _{a})  \label{45}
\end{equation}

We check now that for an optimal $\xi $ in the minimisation program above, $%
\zeta _{a}$ is independent of $a$. Assume $\zeta _{a}>\min_{b}\zeta _{b}$
where the minimum is achieved by some good $b^{\ast }$. We can choose $S$
containing $a$ but not $b^{\ast }$ and such that $\xi _{S}>0$: if this was
impossible $\zeta _{a}\leq \zeta _{b^{\ast }}$ would follow. For $%
\varepsilon $ small enough we construct $\xi ^{\prime }\in \nabla (2^{A})$
identical to $\xi $ except for $\xi _{S}^{\prime }=\xi _{S}-\varepsilon $, $%
\xi _{S\diagdown \{a\}}^{\prime }=\xi _{S\diagdown \{a\}}+\varepsilon $. By
construction the net change in the objective is $-\varepsilon
u(S)+\varepsilon u(S\diagdown \{a\})\leq 0$; moreover $\zeta ^{\prime }$ and 
$\zeta $ coincide everywhere except at $a$ where $\zeta _{a}^{\prime }=\zeta
_{a}-\varepsilon $. We can now choose $\varepsilon $ such that either $\zeta
_{a}^{\prime }=\min_{b}\zeta _{b}$ or $\xi _{S}=0$ and still $\zeta
_{a}^{\prime }>\min_{b}\zeta _{b}$. Then we repeat the construction until
all coordinates of $\zeta $ coincide.

If $\xi $ is deterministic on $\varnothing $ or on $A$, we get the first two
terms in (\ref{32}). For any other $\xi $ we can assume in (\ref{45}) that $%
\xi $ puts no weight on $\varnothing $ or on $A$, and write $\zeta \in
\lbrack 0,1]$ for the common value $\zeta _{a}$. Setting $\delta =\frac{1}{%
\zeta }\xi $ defines a balanced set of weights and $\sum_{T}\xi
_{T}u(T)+x(1-\zeta )=\zeta (\delta \cdot u)+(1-\zeta )x$. Without loss we
can minimise over minimal balanced weights. Finally $\overline{\delta }=%
\frac{1}{\zeta }$ and the proof of (\ref{32}) is complete.

The similar argument for (\ref{34}) starts with

\begin{equation*}
L_{2}(u,x)=\min_{p\in \Delta (x)}\max_{\xi \in \nabla (2^{A})}\sum_{T\in
2^{A}}\xi _{T}(u(T)-p_{T})=\max_{\xi \in \nabla (2^{A})}\{\sum_{T\in
2^{A}}\xi _{T}u(T)-x(\max_{a\in A}\zeta _{a})\}
\end{equation*}%
The critical argument that we can take $\zeta _{a}$ independent of $a$
assumes $\zeta _{a}<\max \zeta =\zeta $, picks $S$ s. t. $\xi _{S}>0$ and
containing $b^{\ast }$ but not $a$ and changes $\xi $ by $\xi _{S}^{\prime
}=\xi _{S}-\varepsilon $, $\xi _{S\cup a}^{\prime }=\xi _{S\cup
a}+\varepsilon $: for $\varepsilon $ small enough the $\max \zeta $ does not
change and the net change on the objective is at least$-\varepsilon
u_{S}+\varepsilon u_{S\cup a}\geq 0$. $\blacksquare \smallskip $

\textbf{Proof of Lemma 4. }Equation (\ref{32}) defines a concave function.
Each term in $x$ increases strictly because $\overline{\delta }>1$ and
reaches $u(A)$ for $x$ large enough, therefore $W_{2}(u;x)$ increases
strictly up to $u(A)$. Similarly in (\ref{34}) $L_{2}(u;x)$ is convex and
strictly decreasing as long as all terms in $x$ are positive, which
terminates for $x$ large enough. So the intersection of $W_{2}(u;\cdot )$
and $L_{2}(u;\cdot )$ as $\Gamma _{2}^{BS}(u|A)$ is well defined.

We proceed now by induction after checking that the function $S\rightarrow
\Gamma _{2}^{BS}(u|S)$ is in $\mathcal{M}^{+}(A)$. Going back to the
definition (\ref{28}) we see that $W_{2}(u;x|S)$ increases weakly in $S$
because agent $u$ can choose in the problem augmented to $S\cup a$ a price
s. t. $p_{a}=0$; and so does\ $L_{2}(u;x|S)$ by (\ref{29}) because in the
augmented problem the agent can choose only among subsets not containing $a$%
. Both $W_{2}(u;x|S)$ and $L_{2}(u;x|S)$ increase weakly in $S$, so their
intersection in $x$ increases too.

The induction step applies Lemma 9 to $\Gamma _{n-1}^{BS}(u|\cdot )\in 
\mathcal{M}^{+}(A)$ and gives $W_{n}(u;x|A),L_{n}(u;x|A)$ by the two programs%
\begin{equation}
W_{n}(u;x|A)=\min \{x,u(A),\min_{\mathcal{B}_{m}}\frac{1}{\overline{\delta }}%
(\delta \cdot \Gamma _{n-1}^{BS}(u|\cdot )-x)+x\}  \label{35}
\end{equation}%
\begin{equation}
L_{n}(u;x|A)=\max \{0,u(A)-(n-1)x,\max_{\mathcal{B}_{m}}\frac{1}{\overline{%
\delta }}(\delta \cdot u-(n-1)x)\}  \label{36}
\end{equation}%
with the properties announced in Lemma 5, and their intersection $\Gamma
_{n}^{BS}(u|\cdot )$ as a function in $\mathcal{M}^{+}(A)$.\smallskip

\textbf{Proof of Proposition 3}. If $u(A)>0$ both functions $%
W_{n}(u;x),L_{n}(u;x)$ are strictly positive for $x$ small enough, proving
Positivity. For Responsiveness we compute formally $\Gamma _{2}^{BS}(u_{F})$
(more rigorously than in the Introduction). First (\ref{32}) gives $%
W_{2}(u_{F};x)=\min \{x,1\}$ because the smallest $\overline{\delta }$ in $%
\mathcal{B}_{m}^{\ast }$ is $\frac{m}{m-1}$ and $L_{2}(u_{F};x)=\max \{0,1-%
\frac{1}{m}x\}$ because the largest $\overline{\delta }$ in $\mathcal{B}%
_{m}^{\ast }$ is $m$. This shows $\Gamma _{2}^{BS}(u_{F}|S)=\frac{|S|}{|S|+1}
$

We omit the straightforward induction argument giving $\Gamma
_{n}^{BS}(u_{F}|S)=\frac{|S|}{|S|+n-1}$.

It remains to check $\Gamma _{n}^{BS}(u)\geq MinMax_{n}(u)$ for all $u$ and $%
n$. This is true for $n=1$. Assume next it holds for $\Gamma _{n-1}^{BS}$
and pick any $u\in \mathcal{M}^{+}(A)$ with optimal bid $x^{\ast }$ where $%
W_{n}$ and intersect. Choose $p\in \Delta (x^{\ast })$ optimal in program (%
\ref{31}) so that $L_{n}(u;x^{\ast }|A)=\max_{\varnothing \subseteq
S\subseteq A}(u(S)-(n-1)p_{S})$. Then (\ref{30}) and the inductive argument
imply%
\begin{equation*}
W_{n}(u;x^{\ast })\geq \min_{\varnothing \subseteq S\subseteq A}(\Gamma
_{n-1}^{BS}(u|S)-p_{S})+x^{\ast }=\Gamma _{n-1}^{BS}(u|T)-p_{T}+x^{\ast }%
\text{ for some }T
\end{equation*}%
\begin{equation*}
\Longrightarrow W_{n}(u;x^{\ast })\geq \frac{1}{n-1}(\overset{n}{u})(\pi
)-p_{T}+x^{\ast }\text{ where }\pi \text{ is some }(n-1)\text{-partition of }%
T
\end{equation*}%
We can now combine this lower bound for $(n-1)W_{n}(u;x^{\ast })$ with $%
L_{n}(u;x^{\ast })\geq u(T^{c})-(n-1)p_{T^{c}}$ to get $n\Gamma
_{n}^{BS}(u)\geq (\overset{n}{u})(\pi )+u(T^{c})$ which completes the proof.

\subsection{Proofs for identical goods}

\textit{We say that the goods }$a,b$\textit{\ are symmetric in }$u$ if we
have%
\begin{equation*}
u(S-b+a)=u(S)\text{ for all }S\text{ s. t. }a\notin S\ni b
\end{equation*}

\textbf{Lemma 10}\textit{\ If two goods }$a,b$\textit{\ are symmetric in }$u$%
\textit{\ their optimal (safe) prices} \textit{in }$W_{n}(x;u|A)$\textit{\
and their worst prices in }$L_{n}(x;u|A)$\textit{\ can be taken equal when
we compute the safe bids in B\&S}$_{n}$\textit{.\smallskip }

\textbf{Proof} For brevity we give the argument for $n=2$ and omit the
obvious induction argument.

Fix $u\in \mathcal{M}^{+}$ and assume $u$ is symmetric in the goods $a,b$.
In the program (\textit{\ref{29}}) defining $L_{2}(u;x)$ assume the\textit{\ 
}worst\textit{\ }price $p$ has $p_{a}<p_{b}$. Let $q$ obtains from $p$ by
averaging $p_{a}$ and $p_{b}$ and changing nothing else. Then $%
\max_{\varnothing \subseteq S\subseteq A}(u(S)-p_{S})$ differs from $%
\max_{\varnothing \subseteq S\subseteq A}(u(S)-q_{S})$ only in pairs of
terms of the form $u(S)-p_{S}$, $u(S-b+a)-p_{S-b+a}$. Replacing $p$ by $q$
lowers the largest of these two terms, so $q$ is still optimal in the
program (\textit{\ref{29}}). The argument for $W_{2}(x;u)$ is identical. $%
\blacksquare $\smallskip

In order to compute now $\Gamma _{2}^{BS}$ when the all the goods are
identical (without assuming convexity or concavity of the utility) we use
the notation\textit{\ }$\partial _{k}u_{\ell }=u_{\ell +k}-u_{\ell }$\textit{%
.\smallskip }

\textbf{Lemma 11 }\textit{Fix a utility }$u$\textit{\ for identical goods.
Agent }$u$\textit{'s optimal bid in the B\&S}$_{2}$\textit{\ rule is}%
\begin{equation}
x^{\ast }=\max \{\frac{m}{m+k}\partial _{k}u_{\ell }|0\leq k,\ell \leq m%
\text{ and }0\leq \ell +k\leq m\}  \label{24}
\end{equation}

\textit{If }$x^{\ast }=\frac{m}{m+k^{\ast }}\partial _{k^{\ast }}u_{\ell
^{\ast }}$\textit{\ his guarantee is}%
\begin{equation}
\Gamma _{2}^{BS}(u)=\frac{\ell ^{\ast }+k^{\ast }}{m+k^{\ast }}u_{\ell
^{\ast }}+\frac{m-\ell ^{\ast }}{m+k^{\ast }}u_{\ell ^{\ast }+k^{\ast }}
\label{25}
\end{equation}

\textbf{Proof }By Lemma 9 in section 11.2 the programs (\ref{28}) and (\ref%
{29}) simplify to%
\begin{equation*}
W_{2}(u;x)=\min_{0\leq s\leq m}\{u_{s}+\frac{m-s}{m}x\}\text{ ; }%
L_{2}(u;x)=\max_{0\leq s\leq m}\{u_{s}-\frac{s}{m}x\}
\end{equation*}

The optimal bid $x^{\ast }$ solves $W_{2}(u;x^{\ast })=L_{2}(u;x^{\ast })$.
Because $W_{2}$ increases and $L_{2}$ decreases, both strictly, the
inequality $x\geq x^{\ast }$ is equivalent to $W_{2}(u;x)\geq L_{2}(u;x)$.
If $s^{\prime }\leq s$ the inequality $u_{s}+\frac{m-s}{m}x\geq u_{s^{\prime
}}-\frac{s^{\prime }}{m}x$ is automatic, therefore $x\geq x^{\ast }$ amounts
to%
\begin{equation*}
u_{\ell }+\frac{m-\ell }{m}x\geq u_{\ell +k}-\frac{\ell +k}{m}x\text{ for
all }k,\ell \geq 0\text{ s. t. }\ell +k\leq m
\end{equation*}%
\begin{equation*}
\Longleftrightarrow x\geq \max_{0\leq \ell +k\leq m}\frac{m}{m+k}(u_{\ell
+k}-u_{\ell })
\end{equation*}%
which proves (\ref{24}) and in turn (\ref{25}). $\blacksquare \smallskip $

Lemma 6 follows at once from this result.\smallskip

\textbf{Proof of statement }$ii)$\textbf{\ in Lemma 7}

For $t\in \lbrack m]$ write $\Gamma _{n}(\theta |t)=\Gamma
_{n}^{BS}(u^{\theta }|T)$ the $n$ person B\&S-guarantee when there are only $%
t$ (identical) goods to distribute. Note that $\Gamma _{n}(\theta |t)=0$ if $%
t<\theta $. We compute first $\Gamma _{2}(\theta |t)$ for $t\geq \theta $:%
\begin{equation*}
W_{2}(\theta ;x|t)=\min \{1,\frac{t-\theta +1}{t}x\}\text{ ; }L_{2}(\theta
;x|t)=\max \{0,1-\frac{t}{m}x\}
\end{equation*}%
\begin{equation*}
\Longrightarrow \Gamma _{2}(\theta |t)=\frac{t-\theta +1}{t+1}\text{ for }%
\theta \leq t\leq m
\end{equation*}%
For $n=3$ equation (\textit{\ref{31}}) is simply: $L_{3}(\theta ;x|m)=\max
\{0,1-\frac{2\theta }{m}x\}$. By the concavity of $t\rightarrow \Gamma
_{2}(\theta |t)$ (\textit{\ref{30}}) becomes%
\begin{equation*}
W_{3}(\theta ;x|m)=\min_{\theta -1\leq t\leq m}\{\Gamma _{2}(\theta |t)+%
\frac{m-t}{m}x\}=\min \{\frac{m-\theta +1}{m}x,\frac{1}{\theta +1}+\frac{%
m-\theta }{m}x\}
\end{equation*}%
after which one checks that the graph of $L_{3}$ intersects that of $W_{3}$
on the line $x\rightarrow \frac{m+1-\theta }{m}x$, and finally $\Gamma
_{3}(\theta |m)=\frac{m+1-\theta }{m+1+\theta }$ with the optimal bid $%
x^{\ast }=\frac{m}{m+1+\theta }$. The general inductive step works in
exactly the same way with the recursive equations (\ref{30}),(\ref{31}).

\subsection{Proof of Proposition 5 statement $i)$}

We assume $n=2$ and omit for brevity the straightforward induction argument
extending the result to any $n$.

Fix any $u\in \mathcal{M}^{+}$; from the proof of Lemma 4 (and Lemma 9 in
that proof) we know that $W_{2}(u;x)$ reaches $u(A)$ at some finite value
denoted $\widetilde{x}(u)$: $W_{2}(u;\cdot )$ increases strictly up to $%
\widetilde{x}(u)$ after which it is flat. Agent $u$'s optimal bid $x^{\ast
}(u)$ in B\&S is strictly below $\widetilde{x}(u)$ (because $W_{2}(u;%
\widetilde{x}(u))=u(A)>L_{2}(u;\widetilde{x}(u)$).

Write now for brevity $\partial _{a}^{+}u=\max_{\varnothing \subseteq
S\subseteq A}\partial _{a}u(S)$ and $\partial _{a}^{-}u=\min_{\varnothing
\subseteq S\subseteq A}\partial _{a}u(S)$.\smallskip

\noindent \textit{Step 1 Fix }$u$\textit{\ and }$x\leq \widetilde{x}(u)$%
\textit{\ and suppose that in the program (\ref{28}) an optimal price is }$%
p\in \Delta (x)$\textit{. Then }$p_{a}\leq \partial _{a}^{+}u$ for all $a$%
\textit{.}

Proof by contradiction: we assume $p_{a}>\partial _{a}^{+}u$ for some $a$
and define a new price $p^{\prime }$ s. t. $p_{a}^{\prime
}=p_{a}-\varepsilon $ and $p_{b}^{\prime }=p_{b}$ otherwise; we choose $%
\varepsilon >0$ small enough that $p_{a}^{\prime }>\partial _{a}^{+}u$ still
holds. For some $T\in 2^{A}$ we have $\min_{\varnothing \subseteq S\subseteq
A}(u(S)-p_{S}^{\prime })=u(T)-p_{T}^{\prime }$. This implies $a\in T$
otherwise adding $a$ to $T$ would contradict the optimality of $T$. We
compute now%
\begin{equation*}
W_{2}(u;x-\varepsilon )\geq u(T)-p_{T}^{\prime }+(x-\varepsilon
)=u(T)-p_{T}+x
\end{equation*}%
\begin{equation*}
\geq \min_{\varnothing \subseteq S\subseteq A}(u(S)-p_{S})+x=W_{2}(u;x)
\end{equation*}%
We see that $W_{2}(u;\cdot )$ is flat before $x$ therefore $x>\widetilde{x}%
(u)$ contradicting the choice of $x$.\smallskip

\noindent \textit{Step 2} Assume $u_{1}$ dominates $u_{2}$ strictly.

A first consequence is $L_{2}(u_{1};x)>L_{2}(u_{2};x)$ for all $x\leq
x^{\ast }(u_{2})$. Indeed $u_{1}(S)-p_{S}>u_{2}(S)-p_{S}$ for all non empty $%
S$ and $p\in \Delta (x)$, and $L_{2}(u_{2};x)$ is positive therefore for any 
$p\in \Delta (x)$ the maximum of $u_{2}(S)-p_{S}$ is achieved at some non
empty $S$.

Next we pick $p\in \Delta (x^{\ast }(u_{2}))$ optimal in (\ref{28}) for $%
u_{2}$. By step 1 and inequality $x^{\ast }(u_{2})<\widetilde{x}(u_{2})$ we
have $p_{a}\leq \partial _{a}^{+}u_{2}<\partial _{a}^{-}u_{1}$ for all $a$,
implying $u_{1}(S)>p_{S}$ for all non empty $S$. Therefore%
\begin{equation*}
W_{2}(u_{1};x^{\ast }(u_{2}))\geq \min_{\varnothing \subseteq S\subseteq
A}(u(S)-p_{S})+x^{\ast }(u_{2})=x^{\ast }(u_{2})
\end{equation*}

Because $W_{2}(u_{1};y)\leq y$ for all $y$ we see that $W_{2}(u_{1};y)=y\geq
W_{2}(u_{2};y)$ for all $y\leq x^{\ast }(u_{2})$.

Gathering the first and last statements in this step we conclude that $%
L_{2}(u_{1};\cdot )$ and $W_{2}(u_{1};\cdot )$ intersect beyond $x^{\ast
}(u_{2})$ so agent $u_{1}$'s safe bid makes her the Seller in stage 2. We
showed a few lines ago $u_{1}(S)>p_{S}$ for any $S$ and any possible price
charged by agent $u_{2}$ therefore agent $u_{1}$ will buy all the goods and
the proof is complete. $\blacksquare $


\begin{thebibliography}{99}
\bibitem{ADG} Alkan A, Demange G, Gale D. 1991. Fair allocation of
indivisible objects and criteria of justice. Econometrica, 59, 1023-1039.

\bibitem{AK} Anand Kumar D. 2025, PhD thesis, Chapter 2, University of
Glasgow.

\bibitem{A} Aragones E. 1995. A derivation of the money rawlsian solution.
Social Choice and Welfare, 12, 3, 267-276

\bibitem{AvKa} Avvakumov S, Karasev R. 2023. Equipartition of a segment.
Maths Operations Research, 48, 1, p. 1-602, C2.

\bibitem{Az} Aziz H. 2020. Developments in multi-agent fair allocation.
AAAI, p. 13563--13568.

\bibitem{Az1} Aziz, H. 2021. Achieving envy-freeness and equitability with
monetary transfers. In Proceedings of the AAAI Conference on Artificial
Intelligence, 35, 6, 5102-5109.

\bibitem{BTF} Babaioff M, Tomer E, Feige U. 2021. Fair-share allocations for
agents with arbitrary entitlements. In Peter Biro, Shuchi Chawla, and
Federico Echenique, editors, Proceedings of the 22d ACM Conference on
Economics and Computation , p.127.

\bibitem{BM1} Bogomolnaia A, Moulin H. 2023. Guarantees in Fair Division:
general and monotone preferences. Maths Operations Research, 48, 1, p.1-602,
C2.

\bibitem{BoLe} Bouveret S, Lemaitre M. 2016. Characterizing conflicts in
fair division of indivisible goods using a scale of criteria. Auton Agent
Multi-Agent Syst, 30:259--290.

\bibitem{BDNSV} Brustle J., Dippel, J., Narayan, V. V., Suzuki, M., Vetta,
A. 2020. One dollar each eliminates envy. In Proceedings of the 21st ACM
Conference on Economics and Computation (pp. 23-39).

\bibitem{Bud} Budish E. 2011. The combinatorial assignment problem:
Approximate competitive equilibrium from equal incomes. \textit{J. Polit.
Econ.} 119, 6, 1061--103.

\bibitem{CI} Caragiannis, I., Ioannidis, S.D. 2022. Computing Envy-Freeable
Allocations with Limited Subsidies. Proceedings of Web and Internet
Economics. WINE 2021. Lecture Notes in Computer Science(), vol 13112.

\bibitem{CGK} Cramton P, Gibbons R, Klemperer P. 1987. Dissolving a
partnership efficiently. Econometrica 55(3):615--632.

\bibitem{CSS} Cramton, P., Shoham, Y., \& Steinberg, R. (Eds.). 2006.
Combinatorial auctions. Cambridge: MIT Press.

\bibitem{C} Crawford V P 1979. A Procedure for Generating Pareto-Efficient
Egalitarian Equivalent Allocations. Econometrica, 47,49-60.

\bibitem{De} Demange G. 1984. Implementing Efficient Egalitarian Equivalent
Allocations, Econometrica, 52,5,1167-1177.

\bibitem{DH} Demko S, Hill T P. 1988. Equitable distribution of indivisible
objects, Math. Soc. Sci., 16, 145-158.

\bibitem{GMPZ} Gal Y, Mash M, Procaccia A D, Zick Y. 2017. Which Is the
Fairest (Rent Division) of Them All?. Journal of the ACM, Vol. 64, No. 6,
Article 39.

\bibitem{GP} Goldman J, Procaccia A D. 2014. Spliddit: Unleashing Fair
Division Algorithms, SIGecom Exchanges 13, 2, 41--46.

\bibitem{Gra} Grabisch M. 2016. Set functions, games and capacities in
decision making. Theory and Decison Library \#46, Springer.

\bibitem{HS} Halpern D., Shah N. 2019. Fair Division with Subsidy. SAGT
2019: 374-389

\bibitem{KK} Kunreuther H, Kleindorfer, P R. 1986. A Sealed-Bid Auction
Mechanism for Siting Noxious Facilities.\ American Economic Review, May 1986
(Papers and Proceedings), 76, 2, 295-299.

\bibitem{Kuh} Kuhn, H.1967. On games of fair division, Essays in
Mathematical Economics in Honour of Oskar Morgenstern, Princeton University
Press, pp. 29--37.

\bibitem{MWG} Mas-Colell A, Whinston MD, Green JR. Microeconomic Theory.
Oxford University Press New York, 1995.

\bibitem{Mo3} Moulin, H. 1992. An application of the Shapley value to fair
division with money. Econometrica, 60, 1331--1349.

\bibitem{Mo4} Moulin H. 2019. Fair Division in the Internet Age, Annual
Review of Economics,11:407-41.

\bibitem{Ni} Nisan N. 2006. Bidding languages. Chapter 1 in \textit{%
Combinatorial Auctions }(MIT Press).

\bibitem{NS} Nisan N, Segal I. 2006. The communication requirements of
efficient allocations and supporting prices. J. Econ. Theory, 129, 1,
192-224.

\bibitem{PR} Plaut B, Roughgarden T. 2020. Communication Complexity of
Discrete Fair Division. SIAM J. Comput. 49(1): 206-243.

\bibitem{PW} Procaccia AD, Wang J. 2014. Fair enough: guaranteeing
approximate maximin shares, Proceedings of the 14th ACM Conference on
Economics and Computation (EC `14), p.675--92. New York: ACM.

\bibitem{St1} Steinhaus H. 1948. The Problem of Fair Division. Econometrica,
16:101--104.

\bibitem{St2} Steinhaus H. 1949. Sur la division pragmatique. Econometrica
(supplement), 17, 315-319.

\bibitem{Sv} Svensson L G. 1983. Large Indivisibles: An Analysis with
Respect to Price Equilibrium and Fairness. Econometrica,51,4,939-954.

\bibitem{TT} Tadenuma K, Thomson W. 1993. The fair allocation of an
indivisible good when monetary compensations are possible. Math Soc Sci 25:
117--132.

\bibitem{W} Walsh T. 2020. Fair division: The computer scientist's
perspective. In IJCAI, pages 4966--4972.
\end{thebibliography}
\end{document}